\documentclass[aps,pre,reprint,letterpaper,showpacs,amsmath,amssymb]{revtex4-1}
\usepackage{graphicx}
\usepackage[absolute]{textpos}
\begin{document}
\begin{textblock*}{8.5in}(0.1in,0.25in)
\begin{center}
PHYSICAL REVIEW E \textbf{88}, 022132 (2013)
\end{center}
\end{textblock*}
\begin{textblock*}{2.5in}(5.6in,10.5in)
\copyright2013 American Physical Society
\end{textblock*}
\title{Bose-Einstein condensation of light: General theory}
\received{7 March 2013}
\published{19 August 2013}
\author{Denis Nikolaevich \surname{Sob'yanin}}
\email{sobyanin@lpi.ru}
\affiliation{Tamm Department of Theoretical Physics,\\Lebedev Physical Institute, Russian Academy of Sciences,\\Leninskii Prospekt 53, Moscow, 119991 Russia}
\begin{abstract}
A theory of Bose-Einstein condensation of light in a dye-filled optical microcavity is presented. The theory is based on the hierarchical maximum entropy principle and allows one to investigate the fluctuating behavior of the photon gas in the microcavity for all numbers of photons, dye molecules, and excitations at all temperatures, including the whole critical region. The master equation describing the interaction between photons and dye molecules in the microcavity is derived and the equivalence between the hierarchical maximum entropy principle and the master equation approach is shown. The cases of a fixed mean total photon number and a fixed total excitation number are considered, and a much sharper, nonparabolic onset of a macroscopic Bose-Einstein condensation of light in the latter case is demonstrated. The theory does not use the grand canonical approximation, takes into account the photon polarization degeneracy, and exactly describes the microscopic, mesoscopic, and macroscopic Bose-Einstein condensation of light. Under certain conditions, it predicts sub-Poissonian statistics of the photon condensate and the polarized photon condensate, and a universal relation takes place between the degrees of second-order coherence for these condensates. In the macroscopic case, there appear a sharp jump in the degrees of second-order coherence, a sharp jump and kink in the reduced standard deviations of the fluctuating numbers of photons in the polarized and whole condensates, and a sharp peak, a cusp, of the Mandel parameter for the whole condensate in the critical region. The possibility of nonclassical light generation in the microcavity with the photon Bose-Einstein condensate is predicted.
\end{abstract}
\maketitle

\section{Introduction}

The aim of this paper is to give a detailed description of the theory of Bose-Einstein condensation (BEC) of light in a dye-filled optical microcavity, a new physical phenomenon that was experimentally observed in 2010~\cite{KlaersEtal2010}. This observation is an important achievement in the study of critical phenomena in optical systems \cite{WeillEtal2005,ConnaughtonEtal2005,ContiEtal2008,NovoaEtal2010,WeillFischerGat2010,WeillEtal2010,SchiroEtal2012,SunEtal2012} and has attracted considerable interest \cite{KlaersEtal2012,Sobyanin2012,ZhangEtal2012a,ZhangEtal2012b,Sobyanin2013a,ZhangEtal2013,SnokeGirvin2013,KruchkovSlyusarenko2013,KirtonKeeling2013,LeeuwStoofDuine2013}. In the experiment, a microcavity composed of two highly reflecting spherical dielectric mirrors is filled with a dye solution and confines photons, which are repeatedly absorbed and reemitted by dye molecules. In the microcavity, we have a photon gas that continuously interacts with the dye solution, and the photon gas can undergo BEC under certain conditions, which reflects in a macroscopic accumulation of photons in the ground cavity mode.

The theory of the light BEC was developed in 2012~\cite{Sobyanin2013a}. It is based on the hierarchical maximum entropy principle~\cite{Sobyanin2012} and describes the microscopic, mesoscopic, and macroscopic BEC of photons at all temperatures, including the whole critical region. The theory presented here generalizes that of Ref.~\cite{Sobyanin2013a} to the case of an arbitrary number of excitations, which can even exceed the number of dye molecules in the microcavity. In addition to the hierarchical maximum entropy principle, an alternative approach to the light BEC---the master equation approach---is developed and the equivalence of the two approaches is demonstrated. Finally, the light BEC is considered not only in the case of a fixed mean total photon number but also in the case of a fixed total excitation number. The theory allows one to fully determine the statistical properties of the photon gas and, in particular, to study fluctuations of different cavity modes, including the condensate and polarized condensate, and fluctuations of the photon gas as a whole. Significantly, the theory does not use the grand canonical approximation and consistently takes into account the photon polarization degeneracy and the interaction of all the cavity modes with dye molecules. The theory predicts the universal relation between the degrees of second-order coherence for the polarized photon condensate and the whole photon condensate. Moreover, it predicts sub-Poissonian photon statistics for these condensates under certain conditions, thereby implying that the photon Bose-Einstein condensate can be used as a new source of nonclassical light. It would be interesting to observe the predictions of the theory experimentally.

This paper is organized as follows: In Sec.~II, the energy spectrum of an optical microcavity is presented, its quantum interpretation is given, and the polarized and whole photon condensates are defined. In Sec.~III, quantum states of the system of photons and dye molecules interacting with each other in the microcavity are described and excitations the number of which is conserved during the interaction are defined. In Sec.~IV, the hierarchical maximum entropy principle is briefly outlined and then applied to the system to investigate its fluctuating behavior; viz., the probability distribution of the system over all quantum states is determined, from which the probability distribution of the photon gas over Fock states and that of the subsystem of dye molecules over the numbers of ground-state and excited molecules are derived. In Sec.~V, the master equation directly describing the interaction between photons and dye molecules in the microcavity is obtained and the equivalence between the hierarchical maximum entropy principle and the master equation approach is shown. In Sec.~VI, fluctuations of the photon gas are investigated in more detail: the probability distributions of the number of photons in the whole photon gas, in each cavity mode, and at each energy level are found and the probability distributions of the number of photons in the condensate and polarized condensate are obtained from the above distributions as a special case. In Sec.~VII, the universal relation between the degrees of second-order coherence for the polarized and whole photon condensates is derived. In Sec.~VIII, an analogy is drawn between the light BEC in an optical microcavity and the usual atomic BEC in a two-dimensional harmonic trap and the critical temperature and temperature dependence of the condensate fraction in the thermodynamic limit are derived from this analogy. In Sec.~IX, the cases of a fixed mean total photon number and a fixed total excitation number are considered, the temperature dependence of the condensate fraction, photon fraction, critical temperature, degrees of second-order coherence, Mandel parameters, and reduced standard deviations for the polarized and whole photon condensates in the microscopic, mesoscopic, and macroscopic cases is discussed, and the possibility of sub-Poissonian photon statistics for the condensates is underlined. In Sec.~X, concluding remarks are given and future prospects are outlined.

\section{Cavity spectrum}

Consider an optical cavity composed of two mirrors with a radius of curvature $R$ that are separated by a distance~$D_0$. The cavity is filled with a dye solution with a refractive index~$n_0$, which determines the speed of light in the medium, $c_0=c/n_0$. The frequency spectrum of the cavity with $D_0<R$ is \cite{BoydGordon1961,BoydKogelnik1962,Vainshtein1963,Vainshtein1966,KogelnikLi1966a,KogelnikLi1966b}
\begin{equation}
\label{freqSpectrumWithArccos}
\frac{\nu}{\nu_\text{f}}=q_0+\frac{m+n+1}{\pi}\arccos\biggl(1-\frac{D_0}{R}\biggr),
\end{equation}
where $\nu$ is a mode frequency, $\nu_\text{f}=c_0/2D_0$ is the cavity free spectral range, the frequency spacing between successive longitudinal resonances, $q_0$ is a longitudinal mode number, equal to the number of half wavelengths fitting into the mirror spacing in the case of longitudinal resonance, and $m$ and $n$ are nonnegative transverse mode numbers. In the experiment~\cite{KlaersEtal2010}, $R\approx1$~m whereas $D_0\approx1.46$~$\mathrm{\mu}$m. Since $D_0\ll R$, we can use in Eq.~\eqref{freqSpectrumWithArccos} the asymptotic formula $\arccos(1-x)\approx\sqrt{2x}$ for a nonnegative $x\ll1$ and write
\begin{equation}
\label{freqSpectrumWithoutArccos}
\hbar\omega=q_0\hbar\omega_\text{f}+(m+n+1)\hbar\Omega,
\end{equation}
where $\omega=2\pi\nu$, $\omega_\text{f}=2\pi\nu_\text{f}$, and
\begin{equation}
\label{Omega}
\Omega=c_0\sqrt{\frac{2}{D_0 R}}.
\end{equation}

Experimentally, the mirror spacing is so small that the free spectral range of the optical microcavity is comparable to the spectral width of the dye. In this case, the emission of photons with a fixed longitudinal number $q_0$ dominates, and the cavity field is a set of $\text{TEM}_{mn}$ modes corresponding to~$q_0$.

Note that the last term on the right-hand side of Eq.~\eqref{freqSpectrumWithoutArccos} coincides with the spectrum of a two-dimensional quantum harmonic oscillator with frequency~$\Omega$. With this observation, we can look at the spectrum~\eqref{freqSpectrumWithoutArccos} from another perspective: In the paraxial approximation~\cite{KlaersEtal2011}, we have
\begin{equation}
\label{biasedEnergyViaH}
\hbar\omega-m_\text{ph}c_0^2=H,
\end{equation}
where
\begin{equation}
\label{Hamiltonian}
H=\frac{p_\text{ph}^2}{2m_\text{ph}}+\frac{m_\text{ph}\Omega^2r^2}{2}
\end{equation}
is the Hamiltonian of a two-dimensional harmonic oscillator with frequency $\Omega$ given by Eq.~\eqref{Omega}, $p_\text{ph}=\hbar k_\perp$ is a transverse photon momentum, $k_\perp$ is a transverse wave number, $r$ is a distance from the optical axis, and
\begin{equation}
\label{photonMass}
m_\text{ph}=\frac{\pi\hbar q_0}{c_0D_0}
\end{equation}
is an effective photon mass formally defined by the relation $m_\text{ph}c_0^2=q_0\hbar\omega_\text{f}$. Thus, the cavity induces a photon mass and provides a harmonic potential, thereby playing the role of a two-dimensional harmonic trap for massive photons; $\Omega$~is the trap frequency.

The quantization procedure for the Hamiltonian~\eqref{Hamiltonian} yields the following eigenvalues:
\begin{equation}
\label{HamiltonianEigenvalues}
\hat{H}=(m+n+1)\hbar\Omega,
\end{equation}
where the nonnegative quantum numbers $m$ and $n$ are exactly the transverse mode numbers entering in Eq.~\eqref{freqSpectrumWithoutArccos}; they are the eigenvalues of Hamiltonians $\hat{H}_x$ and $\hat{H}_y$ for one-dimensional oscillations along two mutually orthogonal directions in the plane orthogonal to the optical axis:
\[
\hat{H}_x=\biggl(m+\frac12\biggr)\hbar\Omega,
\quad
\hat{H}_y=\biggl(n+\frac12\biggr)\hbar\Omega,
\]
with $\hat{H}=\hat{H}_x+\hat{H}_y$. From Eqs.~\eqref{biasedEnergyViaH} and~\eqref{HamiltonianEigenvalues} we get
\begin{equation}
\label{freqSpectrum}
\hbar\omega=\hbar\omega_0+(m+n)\hbar\Omega,
\end{equation}
where
\begin{equation}
\label{groundModeEnergy}
\hbar\omega_0=m_\text{ph}c_0^2+\hbar\Omega
\end{equation}
is the energy of a ground-mode photon, a photon in the TEM$_{00}$ mode. The spectrum~\eqref{freqSpectrum} coincides with that given by Eq.~\eqref{freqSpectrumWithoutArccos}.

In addition to the two nonnegative transverse quantum numbers $m$ and~$n$, there exists a polarization quantum number $p$ that takes on the value $0$ or~$1$ and denotes one of the two photon polarizations. It follows from Eq.~\eqref{freqSpectrum} that the microcavity has a discrete equidistant energy spectrum such that the energy and degeneracy of the $k$th level are
\begin{equation}
\label{finalSpectrum}
\hbar\omega_k=\hbar\omega_0+k\hbar\Omega
\end{equation}
and
\begin{equation}
\label{degeneracy}
g_k=2(k+1),
\end{equation}
respectively; $2$ in Eq.~\eqref{degeneracy} reflects the twofold photon polarization degeneracy.

A photon in the TEM$_{mn}$ mode and with polarization~$p$, which will be called a mode-$mnp$ photon, has energy $\hbar\omega_{m+n}$. A set of mode-$mnp$ photons constitute mode~$mnp$. It is convenient to somehow number all the modes with the same energy. We can then characterize every mode by two quantum numbers: the number~$k$ of the energy level and the number~$i$ of the mode. A possible way to do this is to arrange all the modes $mnp$ with a fixed $k=m+n$ in increasing order of $m$ and then each two modes with a fixed $m$ in increasing order of~$p$, so that mode~$mnp$ has number~$i=2m+p$. Thus, we will deal with mode~$ki$ instead of mode~$mnp$, and a mode-$mnp$ photon becomes a mode-$ki$ photon, with $k=m+n$ and $i=2m+p$. For any given~$k$, the mode number lies in the range
\begin{equation}
\label{iRange}
0\leqslant i\leqslant g_k-1.
\end{equation}

The photon condensate is a set of photons with minimum energy. These condensate photons are ground-mode photons, which have energy $\hbar\omega_0$~\eqref{groundModeEnergy}. The ground state is degenerate and consists of photons of two polarizations. In this connection, we additionally define a polarized ground-mode photon, a ground-mode photon of a definite polarization, and consider a polarized photon condensate, a set of polarized ground-mode photons. The whole photon condensate is then a mixture of two polarized photon condensates, the first comprising mode-$00$ photons and the second comprising mode-$01$ photons. The two polarized condensates obviously have the same statistical properties, but as we will see below, the properties of a polarized condensate differ significantly from those of the condensate considered as a whole. Incidentally, we may mention an analogy with BEC of quasiequilibrium magnons experimentally observed in thin yttrium-iron-garnet films under microwave pumping at room temperature \cite{DemokritovEtal2006,ChumakEtal2009}: the lowest-energy magnon state is twofold degenerate, and the magnon condensate consists of two components with the same frequency but with opposite wavenumbers~\cite{MalomedEtal2010}.

\section{System states}

We can consider the microcavity with photons and dye molecules as composed of two interacting subsystems.

The first subsystem is the photon gas. To characterize its state, consider a set $\{n_{ki}\}$ of the numbers of photons in all cavity modes, with $n_{ki}$ being the number of mode-$ki$ photons. Were all these photon numbers constant, we would have the photon gas in a Fock state~$|\text{ph}\rangle$, which is given by a vector composed of the numbers of photons in each mode:
\[
|\text{ph}\rangle=|\{n_{ki}\}\rangle.
\]
In general, these numbers are not constant and fluctuate due to repeated absorption and reemission of photons by dye molecules. We may thus find the photon gas in the Fock state $|\text{ph}\rangle$ only with a probability~$\pi_{|\text{ph}\rangle}$. To investigate statistical properties of the photon gas and describe its BEC, we should derive the probability distribution~$\{\pi_{|\text{ph}\rangle}\}$.

The second subsystem is the dye solution, which consists of a solvent and a number of dye molecules, $n_\text{d}$. Some dye molecules are in the ground singlet electronic state~$S_0$, the rest are in the first excited singlet electronic state~$S_1$. Let us enumerate all the dye molecules. We can then characterize the state of the subsystem of dye molecules by an $n_\text{d}$-dimensional vector
\[
|\text{d}\rangle=|\{d_j\}\rangle,
\]
where $d_j$, $1\leqslant j\leqslant n_\text{d}$, is $0$ or $1$ if the $j$th dye molecule is, respectively, in the ground or first excited singlet electronic state. If the number of dye molecules of the first kind is $n^0_{|\text{d}\rangle}$ and the number of dye molecules of the second kind is $n^1_{|\text{d}\rangle}$, then
\begin{equation}
\label{dyeMoleculeNumber}
n_\text{d}=n^0_{|\text{d}\rangle}+n^1_{|\text{d}\rangle},
\end{equation}
with
\begin{equation}
\label{excitedMoleculeNumberViaD}
n^1_{|\text{d}\rangle}=\sum_{j=1}^{n_\text{d}}d_j.
\end{equation}

Consider the whole system. Its state is determined by the states of the subsystems and hence is characterized by a joint state vector
\[
|\text{d},\text{ph}\rangle=|\{d_j\},\{n_{ki}\}\rangle.
\]
Because of interaction between the subsystems, this vector fluctuates and has a probability distribution $\{\pi_{|\text{d},\text{ph}\rangle}\}$. A ground-state molecule can absorb a photon and become an excited molecule; conversely, an excited molecule can emit a photon and become a ground-state molecule. Denote by $n_{|\text{ph}\rangle}$ the total photon number:
\begin{equation}
\label{totalPhotonNumber}
n_{|\text{ph}\rangle}=\sum_{k=0}^\infty n_k,
\end{equation}
where
\begin{equation}
\label{PhotonNumberAtEnergyLevel}
n_k=\sum_{i=0}^{2k+1}n_{ki}
\end{equation}
is the number of photons at the $k$th energy level. We can conveniently define excitations in the microcavity as photons and excited dye molecules. The number of excitations so defined is
\begin{equation}
\label{excitationNumber}
n_\Sigma=n^1_{|\text{d}\rangle}+n_{|\text{ph}\rangle}
\end{equation}
and does not change during photon absorption and emission. In this sense, the photon gas and dye solution interact exchanging excitations in a  number-conserving way.

\section{Hierarchical maximum entropy}

\subsection{Premises}

The microcavity is at a temperature~$T$. A significant feature is fast thermalization of the population of the electronic states of dye molecules. Every dye molecule is in contact with the solvent, which plays the role of thermostat of temperature~$T$. The typical thermalization time is $\sim1$~ps at room temperature, the temperature in the experiment~\cite{KlaersEtal2010}, and is short compared with the typical fluorescence lifetime $\sim1{-}10$~ns \cite{Shank1975,SchaeferWillis1976,HaasRotter1991,Schaefer1990,Lakowicz2006}. We thus have apparent separation of the time scales corresponding to thermalization and fluorescence, so that photon emission occurs from thermally equilibrated excited states.

The aforementioned fast thermalization and time-scale separation allow us to consider the microcavity with photons and dye molecules interacting with each other as a generalized superstatistical system~\cite{Sobyanin2011}. Such systems arise in the context of generalized superstatistics, a ``statistics of superstatistics,'' and, as well as ordinary superstatistical systems~\cite{BeckCohen2003}, are characterized by hierarchical structures of dynamics. The hierarchy is formed by the decomposition of the system dynamics into different dynamics on different spatiotemporal scales largely separated from each other. Different fluctuating parameters correspond to different levels of dynamics of a generalized superstatistical system. In our system, we have fluctuations of the energy of dye molecules and those of the state~$|\text{d},\text{ph}\rangle$ of the whole system. The former fluctuations correspond to fast thermalization and are wholly determined by the inverse temperature
\begin{equation}
\label{inverseTemperature}
\beta=(k_\text{B}T)^{-1}.
\end{equation}
The latter fluctuations correspond to the interaction between the photon gas and dye solution, which reflects in the repeated change in the numbers of ground-state and excited dye molecules and photons in each cavity mode. The inverse temperature \eqref{inverseTemperature} in itself does not fluctuate.

The probability distributions of the fluctuating parameters of a generalized superstatistical system can be found using the hierarchical maximum entropy principle~\cite{Sobyanin2012}. This principle generalizes the maximum entropy approaches related to superstatistics~\cite{Crooks2007,AbeBeckCohen2007,StraetenBeck2008,Abe2009,Abe2010} and consists in arranging the levels of dynamics in increasing order of dynamical time scale and consecutively maximizing the Boltzmann-Gibbs-Shannon entropy at each level. In the context of the considered system, the hierarchical maximum entropy principle is reduced to consideration of thermalization of dye molecules, which corresponds to maximization of the entropy at the lower dynamical level, and subsequent maximization of the entropy of the whole system.

\subsection{Thermalization}

Consider thermalization of dye molecules. Let $g_0(\varepsilon)$ and $g_1(\varepsilon)$ be the density of rovibrational states for the ground, $S_0$, and first excited, $S_1$, singlet electronic state, respectively. When defining the density~$g_i(\varepsilon)$, with $i=0$ or~$1$, it is convenient to measure the molecular energy $E$ from the energy of the lowest-energy substate, $E_i$, corresponding to the state~$S_i$: $\varepsilon=E-E_i$. The densities $g_0(\varepsilon)$ and $g_1(\varepsilon)$ are zero for any negative~$\varepsilon$ and become positive as $\varepsilon$ passes through zero, so that $g_i(0)$ is the density of energy states near~$E_i$. Because of thermalization, we have for every dye molecule the Gibbs canonical distribution
\begin{equation}
\label{GibbsDistribution}
\rho_i(E)=\frac{e^{-\beta E}}{Z_i},
\end{equation}
where $\beta$ is the inverse temperature~\eqref{inverseTemperature},
\begin{equation}
\label{partitionFunction}
Z_i=e^{-\beta E_i}z_i
\end{equation}
is the partition function, and
\begin{equation}
\label{reducedPartitionFunction}
z_i=\int_0^\infty e^{-\beta\varepsilon}g_i(\varepsilon)d\varepsilon
\end{equation}
is the reduced partition function, with $i=0$ for a ground-state dye molecule and $1$ for an excited dye molecule. Note that the distribution \eqref{GibbsDistribution} is easily obtained by maximizing the entropy of a dye molecule,
\begin{equation}
\label{generalMoleculeEntropy}
s_i=-\int_{E_i}^\infty\rho_i(E)\ln\rho_i(E)\,g_i(E-E_i)dE,
\end{equation}
under the normalization condition and mean energy constraint~\cite{Sobyanin2012}. From Eqs.~\eqref{GibbsDistribution}--\eqref{generalMoleculeEntropy} we obtain the entropy
\begin{equation}
\label{moleculeEntropy}
s_i=\ln z_i-\frac{\beta}{z_i}\frac{d z_i}{d\beta}
\end{equation}
and mean energy
\begin{equation}
\label{moleculeEnergy}
u_i=E_i-\frac{1}{z_i}\frac{d z_i}{d\beta}
\end{equation}
of a dye molecule.

Equations~\eqref{moleculeEntropy} and~\eqref{moleculeEnergy} allow us to write the entropy
\begin{equation}
\label{dyeSubsystemEntropy}
s_{|\text{d}\rangle}=n^0_{|\text{d}\rangle}s_0+n^1_{|\text{d}\rangle}s_1
\end{equation}
and mean energy
\begin{equation}
\label{dyeSubsystemEnergy}
u_{|\text{d}\rangle}=n^0_{|\text{d}\rangle}u_0+n^1_{|\text{d}\rangle}u_1
\end{equation}
of the subsystem of all the dye molecules in a fixed state~$|\text{d}\rangle$.

The entropy of the photon gas in a fixed Fock state $|\text{ph}\rangle$ is zero,
\begin{equation}
\label{photonGasEntropy}
s_{|\text{ph}\rangle}=0,
\end{equation}
and the corresponding energy is
\begin{equation}
\label{photonGasEnergy}
u_{|\text{ph}\rangle}=n_{|\text{ph}\rangle}\hbar\omega_0+u^\perp_{|\text{ph}\rangle},
\end{equation}
where $u^\perp_{|\text{ph}\rangle}=\hbar\Omega\varepsilon_{|\text{ph}\rangle}$ is the transverse energy and
\begin{equation}
\label{reducedTransverseEnergy}
\varepsilon_{|\text{ph}\rangle}=\sum_{k=0}^\infty k n_k
\end{equation}
is the reduced transverse energy, with $n_k$ defined by Eq.~\eqref{PhotonNumberAtEnergyLevel}. When deriving the energy $u_{|\text{ph}\rangle}$  of the photon gas, we use the cavity spectrum \eqref{finalSpectrum}.

\subsection{General fluctuations}

\subsubsection{Whole system}

Now consider dynamics of the system as a whole. We can express the entropy and mean energy of the whole system in a fixed state~$|\text{d},\text{ph}\rangle$ via the quantities \eqref{dyeSubsystemEntropy}-\eqref{photonGasEnergy}:
\begin{eqnarray}
\label{wholeSystemEntropyFixedState}
S_{|\text{d},\text{ph}\rangle}&=&s_{|\text{d}\rangle}+s_{|\text{ph}\rangle},
\\
\nonumber
U_{|\text{d},\text{ph}\rangle}&=&u_{|\text{d}\rangle}+u_{|\text{ph}\rangle}.
\end{eqnarray}
However, $|\text{d},\text{ph}\rangle$ fluctuates, and the entropy of the whole system is given by
\begin{equation}
\label{wholeSystemEntropy}
S=-\sum_{|\text{d},\text{ph}\rangle}\pi_{|\text{d},\text{ph}\rangle}\ln\pi_{|\text{d},\text{ph}\rangle}
+\sum_{|\text{d},\text{ph}\rangle}\pi_{|\text{d},\text{ph}\rangle}S_{|\text{d},\text{ph}\rangle}.
\end{equation}
Writing $\sum_{|\text{d},\text{ph}\rangle}$ implies summation over all states $|\text{d},\text{ph}\rangle$ such that the total numbers of dye molecules and excitations are fixed and that the numbers of ground-state dye molecules, excited dye molecules, and photons in different cavity modes are nonnegative and subject to Eqs.~\eqref{dyeMoleculeNumber} and~\eqref{totalPhotonNumber}--\eqref{excitationNumber} but otherwise arbitrary. To find the probability distribution~$\{\pi_{|\text{d},\text{ph}\rangle}\}$, we should maximize the entropy~\eqref{wholeSystemEntropy}
under the normalization condition
\begin{equation}
\label{normalizationCondition}
\sum_{|\text{d},\text{ph}\rangle}\pi_{|\text{d},\text{ph}\rangle}=1
\end{equation}
and mean energy constraint
\[
\sum_{|\text{d},\text{ph}\rangle}\pi_{|\text{d},\text{ph}\rangle}U_{|\text{d},\text{ph}\rangle}=U.
\]
The condition of zero variation, $\delta L=0$, for the Lagrange function
\[
L=S-(\nu-1)\sum_{|\text{d},\text{ph}\rangle}\pi_{|\text{d},\text{ph}\rangle}-\beta\sum_{|\text{d},\text{ph}\rangle}\pi_{|\text{d},\text{ph}\rangle}U_{|\text{d},\text{ph}\rangle}
\]
yields the probability of state $|\text{d},\text{ph}\rangle$:
\begin{equation}
\label{dPhProbability}
\pi_{|\text{d},\text{ph}\rangle}=\frac{z_0^{n^0_{|\text{d}\rangle}}z_1^{n^1_{|\text{d}\rangle}}}{Z}\exp[-\beta(n_\text{d} E_0+n^1_{|\text{d}\rangle}\hbar\omega_\text{e}+u_{|\text{ph}\rangle})],
\end{equation}
where $Z=e^\nu$ is the partition function determined from the normalization condition~\eqref{normalizationCondition},
\begin{equation}
\label{Z}
Z=\sum_{|\text{d},\text{ph}\rangle}z_0^{n^0_{|\text{d}\rangle}}z_1^{n^1_{|\text{d}\rangle}}\exp[-\beta(n_\text{d} E_0+n^1_{|\text{d}\rangle}\hbar\omega_\text{e}+u_{|\text{ph}\rangle})],
\end{equation}
\begin{equation}
\label{energySpacing}
\hbar\omega_\text{e}=E_1-E_0
\end{equation}
is the energy spacing between the lowest-energy substates of~$S_0$ and~$S_1$, and the photon gas energy $u_{|\text{ph}\rangle}$ is given by Eq.~\eqref{photonGasEnergy}.

The probability $\pi_{|\text{d},\text{ph}\rangle}$ is determined by Eq.~\eqref{dPhProbability} only for the states $|\text{d},\text{ph}\rangle$ described after Eq.~\eqref{wholeSystemEntropy}. For all the other states, $\pi_{|\text{d},\text{ph}\rangle}\equiv0$. In this connection, we should note that, by Eqs.~\eqref{dyeMoleculeNumber} and~\eqref{excitationNumber}, the numbers of ground-state and excited dye molecules are in fact functions of Fock state~$|\text{ph}\rangle$, given $n_\text{d}$ and~$n_\Sigma$:
\begin{eqnarray}
\label{groundMoleculeNumberOnFockState}
n^0_{|\text{d}\rangle}&\equiv& n^0_{|\text{ph}\rangle}=n_\text{d}-n_\Sigma+n_{|\text{ph}\rangle},
\\
\label{excitedMoleculeNumberOnFockState}
n^1_{|\text{d}\rangle}&\equiv& n^1_{|\text{ph}\rangle}=n_\Sigma-n_{|\text{ph}\rangle}.
\end{eqnarray}
Therefore, $\pi_{|\text{d},\text{ph}\rangle}$ is a function of~$|\text{ph}\rangle$,
\begin{equation}
\label{piDPhOnPh}
\pi_{|\text{d},\text{ph}\rangle}\equiv\tilde{\pi}_{|\text{ph}\rangle},
\end{equation}
where, by Eqs.~\eqref{dPhProbability}, \eqref{groundMoleculeNumberOnFockState}, and~\eqref{excitedMoleculeNumberOnFockState},
\begin{eqnarray}
\tilde{\pi}_{|\text{ph}\rangle}&=&\frac{z_0^{n_\text{d}-n_\Sigma+n_{|\text{ph}\rangle}}z_1^{n_\Sigma-n_{|\text{ph}\rangle}}}{Z}
\label{piTilde}
\\
& &\times\exp\{-\beta[n_\text{d} E_0+(n_\Sigma-n_{|\text{ph}\rangle})\hbar\omega_\text{e}+u_{|\text{ph}\rangle}]\}.
\nonumber
\end{eqnarray}

\subsubsection{Photon gas}

Let $|\text{d}|\text{ph}\rangle$ be a state of the subsystem of dye molecules given a fixed Fock state of the photon gas, $|\text{ph}\rangle$. The distribution $\{\pi_{|\text{ph}\rangle}\}$ is obtained by summing $\pi_{|\text{d},\text{ph}\rangle}$ over all states $|\text{d}|\text{ph}\rangle$:
\begin{equation}
\label{sumOfPiDPhOverDGivenPh}
\pi_{|\text{ph}\rangle}=\sum_{|\text{d}|\text{ph}\rangle}\pi_{|\text{d},\text{ph}\rangle}.
\end{equation}
Summation not over all $2^{n_\text{d}}$ states $|\text{d}\rangle$ is due to the fact that given $|\text{ph}\rangle$, not all states can be realized because of the conservation of the total excitation number~$n_\Sigma$ [see Eq.~\eqref{excitationNumber}]. Equation~\eqref{piDPhOnPh} allows us to factor out $\pi_{|\text{d},\text{ph}\rangle}$ from the sum:
\begin{equation}
\label{piPhViaPiTilde}
\pi_{|\text{ph}\rangle}=\tilde{\pi}_{|\text{ph}\rangle}W_{|\text{ph}\rangle},
\end{equation}
where the number of states $|\text{d}|\text{ph}\rangle$,
\[
W_{|\text{ph}\rangle}=\sum_{|\text{d}|\text{ph}\rangle}1,
\]
is given by the binomial coefficient ${n_\text{d}\choose n^1_{|\text{d}\rangle}}$, which is the number of $n_\text{d}$-digit binary numbers $d_1d_2\cdots d_{n_\text{d}}$ with exactly $n^1_{|\text{d}\rangle}$ unities; using Eq.~\eqref{excitedMoleculeNumberOnFockState} gives
\begin{equation}
\label{sumOfStatesDGivenPh}
W_{|\text{ph}\rangle}=
{n_\text{d}\choose n_\Sigma-n_{|\text{ph}\rangle}}.
\end{equation}

With Eqs.~\eqref{piTilde}, \eqref{piPhViaPiTilde}, and~\eqref{sumOfStatesDGivenPh}, we obtain the probability distribution of the photon gas over Fock states, $\{\pi_{|\text{ph}\rangle}\}$, which is the main result of the light BEC theory:
\begin{equation}
\label{mainPiPh}
\pi_{|\text{ph}\rangle}=P
{n_\text{d}\choose n_\Sigma-n_{|\text{ph}\rangle}}
r^{n_{|\text{ph}\rangle}}q^{\varepsilon_{|\text{ph}\rangle}},
\end{equation}
where
\begin{gather}
P=\frac{z_0^{n_\text{d}-n_\Sigma}z_1^{n_\Sigma}}{Z}\exp[-\beta(n_\text{d}E_0+n_\Sigma\hbar\omega_\text{e})],
\nonumber\\
\label{r}
r=\frac{z_0}{z_1}\exp[-\beta\hbar(\omega_0-\omega_\text{e})],
\\
\label{q}
q=\exp(-\beta\hbar\Omega),
\end{gather}
the partition function $Z$ is given by Eq.~\eqref{Z}, reduced partition functions $z_0$ and $z_1$ by Eq.~\eqref{reducedPartitionFunction}, inverse temperature $\beta$ by Eq.~\eqref{inverseTemperature}, energy spacing $\hbar\omega_\text{e}$ by Eq.~\eqref{energySpacing}, ground-mode photon energy $\hbar\omega_0$ by Eq.~\eqref{groundModeEnergy}, trap frequency $\Omega$ by Eq.~\eqref{Omega}, total photon number $n_{|\text{ph}\rangle}$ by Eq.~\eqref{totalPhotonNumber}, and reduced transverse energy $\varepsilon_{|\text{ph}\rangle}$ by Eq.~\eqref{reducedTransverseEnergy}. For any finite $T\geqslant0$,
\[
0\leqslant q<1.
\]
The normalization coefficient $P$ can also be found directly from the normalization condition
\[
\sum_{|\text{ph}\rangle}\pi_{|\text{ph}\rangle}=1
\]
for the probability distribution $\{\pi_{|\text{ph}\rangle}\}$:
\begin{equation}
\label{P1}
P=\biggl[\,
\sum_{|\text{ph}\rangle}
{n_\text{d}\choose n_\Sigma-n_{|\text{ph}\rangle}}
r^{n_{|\text{ph}\rangle}}q^{\varepsilon_{|\text{ph}\rangle}}
\biggr]^{-1}.
\end{equation}

Recall that the numbers of photons and excited dye molecules are not independent but subject to the condition~\eqref{excitationNumber}. If all the dye molecules are in the ground state, then all the excitations are photons. Therefore, the maximum photon number equals~$n_\Sigma$. If the number of excitations does not exceed the number of dye molecules, $n_\Sigma\leqslant n_\text{d}$, then all $n_\Sigma$ excitations can be excited dye molecules and the corresponding minimum photon number is~$0$. If we have the opposite situation, $n_\Sigma>n_\text{d}$, then the minimum photon number is positive and equals $n_\Sigma-n_\text{d}$ because at most $n_\text{d}$ dye molecules can be in the excited state. The probability distribution $\{\pi_{|\text{ph}\rangle}\}$ is thus determined by Eq.~\eqref{mainPiPh} for all states $|\text{ph}\rangle$ such that the total photon number $n_{|\text{ph}\rangle}$ lies in the range
\begin{equation}
\label{photonNumberRange}
n^\text{ph}_\text{min}\leqslant n_{|\text{ph}\rangle}\leqslant n_\Sigma,
\end{equation}
where the minimum photon number is
\begin{equation}
\label{nPhMin}
n^\text{ph}_\text{min}=\max\{0,n_\Sigma-n_\text{d}\}.
\end{equation}
Summation in Eq.~\eqref{P1} is performed only over these Fock states. For all the other states, $\pi_{|\text{ph}\rangle}\equiv0$.

\subsubsection{Dye solution}

For completeness, consider fluctuations of the subsystem of dye molecules. The conditional probability of state $|\text{d}\rangle$ given $|\text{ph}\rangle$ is by definition
\begin{equation}
\label{piDGivenPhDefinition}
\pi_{|\text{d}|\text{ph}\rangle}=\frac{\pi_{|\text{d},\text{ph}\rangle}}{\pi_{|\text{ph}\rangle}}.
\end{equation}
Substituting Eq.~\eqref{piDPhOnPh} in Eq.~\eqref{piDGivenPhDefinition} and comparing the latter to Eq.~\eqref{piPhViaPiTilde}, we have
\begin{equation}
\label{finalPiDGivenPhDefinition}
\pi_{|\text{d}|\text{ph}\rangle}=W_{|\text{ph}\rangle}^{-1},
\end{equation}
where $W_{|\text{ph}\rangle}$ is given by Eq.~\eqref{sumOfStatesDGivenPh}. We see that all the states $|\text{d}|\text{ph}\rangle$ are equiprobable.

In this connection, an alternative derivation of $\{\pi_{|\text{ph}\rangle}\}$ can be given, viz., the fluctuating Fock state $|\text{ph}\rangle$ can be used as the parameter describing interaction between the two subsystems. This approach was used in Ref.~\cite{Sobyanin2013a}. In this case, $|\text{ph}\rangle$ replaces $|\text{d},\text{ph}\rangle$ in the expression \eqref{wholeSystemEntropy} for the entropy $S$ of the whole system, so that $S$ directly includes the probability distribution of the photon gas, $\{\pi_{|\text{ph}\rangle}\}$, instead of that of the whole system, $\{\pi_{|\text{d},\text{ph}\rangle}\}$. In addition, $S_{|\text{d},\text{ph}\rangle}$ is replaced by the entropy $S_{|\text{ph}\rangle}$ of the whole system given $|\text{ph}\rangle$,
\begin{equation}
\label{wholeSystemEntropyOnPh}
S_{|\text{ph}\rangle}=\tilde{S}_{|\text{ph}\rangle}+\ln W_{|\text{ph}\rangle},
\end{equation}
where $\tilde{S}_{|\text{ph}\rangle}$ is the entropy $S_{|\text{d},\text{ph}\rangle}$ considered as a function of~$|\text{ph}\rangle$ [see Eqs.~\eqref{dyeSubsystemEntropy}, \eqref{photonGasEntropy}, \eqref{wholeSystemEntropyFixedState}, \eqref{groundMoleculeNumberOnFockState}, and~\eqref{excitedMoleculeNumberOnFockState}],
\[
\tilde{S}_{|\text{ph}\rangle}\equiv S_{|\text{d},\text{ph}\rangle}=
(n_\text{d}-n_\Sigma+n_{|\text{ph}\rangle})s_0+(n_\Sigma-n_{|\text{ph}\rangle})s_1.
\]
The entropy of the whole system corresponding to ${|\text{ph}\rangle}$ is greater than the same entropy corresponding to ${|\text{d},\text{ph}\rangle}$ because the Fock state ${|\text{ph}\rangle}$ of the photon gas describes the system in less detail than the state ${|\text{d},\text{ph}\rangle}$ of the whole system. The additional entropy is exactly the logarithm of the binomial coefficient $W_{|\text{ph}\rangle}$~\eqref{sumOfStatesDGivenPh}. This fact reflects the Boltzmann principle, and $W_{|\text{ph}\rangle}$ is the statistical weight equal to the number of microstates $|\text{d}|\text{ph}\rangle$ corresponding to the Fock macrostate~${|\text{ph}\rangle}$.

In the approach of Ref.~\cite{Sobyanin2013a}, Eq.~\eqref{wholeSystemEntropyOnPh} is obtained using obvious equiprobability of the distribution~$\{\pi_{|\text{d}|\text{ph}\rangle}\}$. The approach described in the present paper is more complex than that used in Ref.~\cite{Sobyanin2013a} because of additionally considering a more detailed distribution $\{\pi_{|\text{d},\text{ph}\rangle}\}$ and subsequently summing over states $|\text{d}|\text{ph}\rangle$ with Eq.~\eqref{sumOfPiDPhOverDGivenPh} to obtain the probability distribution~$\{\pi_{|\text{ph}\rangle}\}$, which is of primary interest. However, it is more general in the sense that equiprobability of~$\{\pi_{|\text{d}|\text{ph}\rangle}\}$ [Eq.~\eqref{finalPiDGivenPhDefinition}] is derived directly from the hierarchical maximum entropy principle. Both the approaches give the same results.

Let $|\text{ph}|\text{d}\rangle$ be a Fock state of the photon gas given a fixed state $|\text{d}\rangle$ of the subsystem of dye molecules. The probability $\pi_{|\text{d}\rangle}$ that the subsystem is in the state $|\text{d}\rangle$ is
\begin{equation}
\label{piD}
\pi_{|\text{d}\rangle}=\sum_{|\text{ph}|\text{d}\rangle}\pi_{|\text{d},\text{ph}\rangle}.
\end{equation}
Choosing $|\text{d}\rangle$ fixes the total photon number: $n_{|\text{ph}\rangle}=n_\Sigma-n^1_{|\text{d}\rangle}$, with $n^1_{|\text{d}\rangle}$ given by Eq.~\eqref{excitedMoleculeNumberViaD}. Denote by $|\text{ph}^n\rangle$ a Fock state with $n$ photons: $n_{|\text{ph}^n\rangle}=n$. Using Eq.~\eqref{dPhProbability}, we obtain from Eq.~\eqref{piD}
\begin{equation}
\label{piD1}
\pi_{|\text{d}\rangle}=Q\,r^{-n^1_{|\text{d}\rangle}}a_{n_\Sigma-n^1_{|\text{d}\rangle}},
\end{equation}
where
\[
Q=\frac{z_0^{n_\text{d}}}{Z}\exp[-\beta(n_\text{d}E_0+n_\Sigma\hbar\omega_0)]
\]
and
\begin{equation}
\label{aN}
a_n=\sum_{|\text{ph}^n\rangle}q^{\varepsilon_{|\text{ph}\rangle}},\quad n\geqslant0.
\end{equation}

Every excited dye molecule can emit a photon and make a transition to the ground state. Therefore, the minimum number of excited dye molecules in the dye solution is~$0$. The maximum number of excited dye molecules is bounded above by the total number of dye molecules, $n_\text{d}$, but at the same time cannot exceed the total excitation number $n_\Sigma$ if $n_\Sigma\leqslant n_\text{d}$. The probability distribution $\{\pi_{|\text{d}\rangle}\}$ is thus determined by Eq.~\eqref{piD1} for all states $|\text{d}\rangle$ such that the number of excited dye molecules lies in the range
\begin{equation}
\label{excitedMoleculeNumberRange}
0\leqslant n^1_{|\text{d}\rangle}\leqslant n^\text{e}_\text{max},
\end{equation}
where the maximum number of excited dye molecules is
\[
n^\text{e}_\text{max}=\min\{n_\Sigma,n_\text{d}\}.
\]
For all the other states, $\pi_{|\text{d}\rangle}\equiv0$.

The normalization coefficient $Q$ can also be determined directly from the normalization condition
\begin{equation}
\label{piDNormalization}
\sum_{|\text{d}\rangle}\pi_{|\text{d}\rangle}=1.
\end{equation}
We see from Eq.~\eqref{piD1} that $\pi_{|\text{d}\rangle}$ in fact depends on the number of excited dye molecules, ${n^1_{|\text{d}\rangle}}$, which means that all the states $|\text{d}\rangle$ corresponding to the same ${n^1_{|\text{d}\rangle}}$ are equiprobable. We again arrive at the equiprobability condition~\eqref{finalPiDGivenPhDefinition}. This condition makes it convenient in Eq.~\eqref{piDNormalization} first to trivially sum over the states with fixed ${n^1_{|\text{d}\rangle}}$ and then to sum over all possible ${n^1_{|\text{d}\rangle}}$ from the range~\eqref{excitedMoleculeNumberRange}:
\begin{equation}
\label{finalQ}
Q=\biggl[\,\sum_{n=0}^{n^\text{e}_\text{max}}{n_\text{d}\choose n}r^{-n} a_{n_\Sigma-n}\biggr]^{-1}.
\end{equation}

It remains to study the sums~\eqref{aN}. This will be done in Sec.~VI\,A.

\section{Master equation}

In Sec.~IV, the theory of the light BEC was constructed using the hierarchical maximum entropy principle. However, photons and dye molecules interact with each other in the microcavity and a master equation can be written with terms on the right-hand side directly describing this interaction. Therefore, it should be possible to obtain the results of the light BEC theory not only with the hierarchical maximum entropy principle but also with the master equation approach. The question naturally arises about the equivalence of these two approaches. The consideration of a simplified case when the ground mode is coupled to a fixed number of dye molecules and the photon polarization degeneracy is absent argues for the equivalence: Applying the hierarchical maximum entropy principle to the simplified system gives the same results as those obtained from solving the corresponding master equation~\cite{Sobyanin2012}. It is natural to expect the same equivalence when the above simplifications are not used. The aim of this section is to obtain the master equation describing the interaction between photons in all the cavity modes and dye molecules and to show the equivalence of the hierarchical maximum entropy principle and the master equation approach in the general case. Note that the master equation approach was applied to the usual atomic BEC \cite{Scully1999,KocharovskyEtal2000,ScullySvidzinsky2006,SvidzinskyScully2006,SvidzinskyScully2010}, and consideration of this approach in the context of the light BEC is interesting in itself.

Denote by $|\text{ph}\,ki^1\rangle$ the state that differs from the state $|\text{ph}\rangle$ only by the presence of one additional mode-$ki$ photon and by $|\text{ph}\,ki^{-1}\rangle$ the state that differs from the state $|\text{ph}\rangle$ only by the absence of one mode-$ki$ photon. Then we can write the master equation
\begin{widetext}
\begin{eqnarray}
\dot{\pi}_{|\text{ph}\rangle}&=&
(n_\Sigma-n_{|\text{ph}\rangle}+1)\sum_{k=0}^\infty\sum_{i=0}^{2k+1}\pi_{|\text{ph}\,ki^{-1}\rangle}R^{10}_{ki}n_{ki}
+(n_\text{d}-n_\Sigma+n_{|\text{ph}\rangle}+1)\sum_{k=0}^\infty\sum_{i=0}^{2k+1}\pi_{|\text{ph}\,ki^1\rangle}R^{01}_{ki}(n_{ki}+1)
\nonumber\\
& &-(n_\Sigma-n_{|\text{ph}\rangle})\pi_{|\text{ph}\rangle}\sum_{k=0}^\infty\sum_{i=0}^{2k+1}R^{10}_{ki}(n_{ki}+1)
-(n_\text{d}-n_\Sigma+n_{|\text{ph}\rangle})\pi_{|\text{ph}\rangle}\sum_{k=0}^\infty\sum_{i=0}^{2k+1}R^{01}_{ki}n_{ki},
\label{masterEquation}
\end{eqnarray}
\end{widetext}
where $R^{10}_{ki}$ and $R^{01}_{ki}$ are, respectively, the rates (per dye molecule per photon) for stimulated emission and absorption of mode-$ki$ photons. In Eq.~\eqref{masterEquation}, we should formally put $\pi_{|\text{ph}\rangle}\equiv0$ if an $n_{ki}<0$ in $|\text{ph}\rangle$, if $n_{|\text{ph}\rangle}<n^\text{ph}_\text{min}$, or if $n_{|\text{ph}\rangle}>n_\Sigma$. The first and third terms on the right-hand side of Eq.~\eqref{masterEquation} correspond to spontaneous and stimulated photon emission, and the second and fourth terms correspond to photon absorption.

To observe the equivalence between the hierarchical maximum entropy principle and the master equation approach, we need not solve the master equation \eqref{masterEquation} because the probability distribution $\{\pi_{|\text{ph}\rangle}\}$ is already known and given by Eq.~\eqref{mainPiPh}. We should only check that this $\{\pi_{|\text{ph}\rangle}\}$ is the solution of the master equation. However, in contrast to the master equation, $\{\pi_{|\text{ph}\rangle}\}$ does not explicitly contain the stimulated photon emission and absorption rates. Therefore, we may expect that the latter are related to each other. This relation can be obtained by writing
\begin{equation}
\label{rates}
R^{10}_{ki}=\sigma_{10}(\omega_k)\frac{I_{ki}}{\hbar\omega_k},\quad
R^{01}_{ki}=\sigma_{01}(\omega_k)\frac{I_{ki}}{\hbar\omega_k},
\end{equation}
where $I_{ki}$ is an effective intensity of mode $ki$ per photon and $\sigma_{10}(\omega)$ and $\sigma_{01}(\omega)$ are, respectively, the cross sections for stimulated emission and absorption. Under the assumption that the populations of the rovibrational degrees of freedom within both the ground and the first excited electronic states of a dye molecule are thermally equilibrated and are at the same temperature $T$, $\sigma_{10}(\omega)$ and $\sigma_{01}(\omega)$ are related by \cite{BandHeller1988}
\begin{equation}
\label{KSLawViaSigma}
\frac{\sigma_{10}(\omega)}{\sigma_{01}(\omega)}=\frac{z_0}{z_1}\exp[-\beta\hbar(\omega-\omega_\text{e})].
\end{equation}
Note that, for the validity of the relation \eqref{KSLawViaSigma}, complete thermal equilibrium is unnecessary (e.g., there may be a population inversion between the ground and first excited electronic states), and it is sufficient that each electronic state is thermalized separately. Now, substituting Eq.~\eqref{mainPiPh} in Eq.~\eqref{masterEquation} and using Eqs.~\eqref{rates} and~\eqref{KSLawViaSigma}, we can check that the probability distribution $\{\pi_{|\text{ph}\rangle}\}$ determined with the hierarchical maximum entropy principle is the stationary solution of the master equation. In this sense, the hierarchical maximum entropy principle provides an alternative to the master equation approach. In fact, the interaction between photons and dye molecules is the dynamical reason for an increase in the entropy of the system.

An attempt to use the master equation approach in the context of the light BEC was made in Ref.~\cite{KlaersEtal2012}. Using the grand canonical approximation and neglecting the twofold polarization degeneracy, the authors postulate that the photon condensate interacts with a reservoir of $M$ dye molecules so that the sum $X$ of the number of excited dye molecules and ground-mode photons is fixed; at the same time, they assume that photons at higher energy levels have the Bose-Einstein distribution. The authors calculate the average condensate photon number $\bar{n}_0$ as a function of $M$ and~$X$ by solving the master equation describing the interaction between dye molecules and condensate photons. Then the authors calculate the average number of photons in the excited modes, $\bar{n}_\text{exc}$, from the chemical potential, which is in turn determined from the ratio $(X-\bar{n}_0)/(M-X+\bar{n}_0)$, so that $\bar{n}_\text{exc}$ is also a function of $M$ and~$X$. Given $M$, the authors adjust $X$ so as to obtain a predetermined value of the average total photon number $\bar{n}_0+\bar{n}_\text{exc}$.

Unfortunately, the work \cite{KlaersEtal2012} is incorrect. It is unphysical to artificially fix the quantity~$X$, which probabilistically takes on all values from the range $0\leqslant X\leqslant n_\Sigma$ and is characterized by a probability distribution because every ground-state dye molecule can absorb and every excited dye molecule can emit not only a ground-mode photon but also a photon with any energy from the cavity spectrum at a nonzero temperature. This fixing results in the distortion of the correct statistical properties of the photon gas. An additional distortion is due to the grand canonical approximation, which imposes a priori constraints on the photon statistics. Significantly, the master equation proposed in Ref.~\cite{KlaersEtal2012} not only does not take account of the photon polarization degeneracy and the interaction between dye molecules and photons in excited cavity modes---though both the former and the latter phenomena always take place---but also is incorrectly written under the assumptions used in Ref.~\cite{KlaersEtal2012}: The rates of stimulated photon emission and absorption are wrongly assumed to be the products of the corresponding Einstein coefficients and the spectral energy density of the ground mode. The ideal microcavity has zero frequency width and infinite spectral energy density of every cavity mode, and the authors' assumption leads to infinite emission and absorption rates. This result implies infinite intensity of the interaction and, hence, infinite speed of reaching equilibrium. The authors try to sidestep this problem by artificially introducing a finite damping in the ideal microcavity, which results in a finite spectral energy density due to small broadening the mode energy. However, this attempt cannot be considered correct because the formulas for the rates used by the authors are inapplicable to the case of small broadening. Moreover, when calculating the rates, the authors erroneously use the Einstein coefficients that enter the universal Stepanov relation between the absorption and emission spectra of dye molecules \cite{Stepanov1957}. Note that this relation takes place for the spectral Einstein coefficients, so that the dimensions of the left- and right-hand sides of the master equation proposed in Ref.~\cite{KlaersEtal2012} do not coincide.

Thus, the erroneous consideration given in Ref.~\cite{KlaersEtal2012} has no relation to the experimental situation of Ref.~\cite{KlaersEtal2010} and leads the authors to wrong conclusions about the statistical properties of the system; note in this connection an incorrect result of Ref.~\cite{KlaersEtal2012} about the Poisson condensate statistics at low temperatures [cf. Sec.~IX]. The master equation~\eqref{masterEquation} does not contain the mistakes of Ref.~\cite{KlaersEtal2012}, takes into account the photon polarization degeneracy and the interaction of all the cavity modes with dye molecules, does not require using the grand canonical approximation, and is not restricted to the thermodynamic limit.

\section{Photon gas fluctuations}

General fluctuations of the whole system, photon gas, and dye solution have been investigated in Sec.~IV and are given by Eqs.~\eqref{dPhProbability}, \eqref{mainPiPh}, and~\eqref{piD1}, respectively. Since it is the behavior of the photon gas that is primary importance for us, we will consider in this section fluctuations of the photon gas in more detail. Starting from the main formula~\eqref{mainPiPh}, which represents the probability distribution over all Fock states and determines all statistical characteristics of the photon gas, we will derive the probability distributions for the total number of photons, the number of photons in each cavity mode, and the number of photons with a fixed energy and last consider condensate fluctuations.

\subsection{Whole photon gas}

First consider fluctuations of the photon gas as a whole. The probability that the photon gas comprises $n$ photons is obtained by summing $\pi_{|\text{ph}\rangle}$ over all Fock states with $n$ photons, $|\text{ph}^n\rangle$:
\[
\pi_n=\sum_{|\text{ph}^n\rangle}\pi_{|\text{ph}\rangle}.
\]
Hence,
\begin{equation}
\label{piN}
\pi_n=P{n_\text{d}\choose n_\Sigma-n}r^n a_n,
\end{equation}
where $r$ and $a_n$ are given by Eqs.~\eqref{r} and~\eqref{aN}, respectively. Equation~\eqref{piN} allows us to find the normalization coefficient [cf. Eq.~\eqref{P1}]
\begin{equation}
\label{finalP}
P=\biggl[\sum_{n=n^\text{ph}_\text{min}}^{n_\Sigma}{n_\text{d}\choose n_\Sigma-n}r^n a_n\biggr]^{-1}
\end{equation}
and then calculate $\pi_n$ for all possible $n$ from the range [see Eq.~\eqref{photonNumberRange}]
\begin{equation}
\label{photonNumberRange1}
n^\text{ph}_\text{min}\leqslant n\leqslant n_\Sigma,
\end{equation}
where $n^\text{ph}_\text{min}$ is defined by Eq.~\eqref{nPhMin}. For all the other~$n$, $\pi_n\equiv0$.

To find $a_n$, let us use the following combinatorial idea: Associate with each photon mode a power series
\[
\sigma_{ki}(z_{ki})=\kappa^{(0)}_{ki}+\kappa^{(1)}_{ki}z_{ki}+\kappa^{(2)}_{ki}z^2_{ki}+\cdots
\]
and consider the product of these series, $\prod\sigma_{ki}(z_{ki})$. This product represents an infinite sum such that it is possible to associate each term in this sum with a Fock state in a one-to-one manner. Specifically, the term $\prod\kappa_{ki}^{(n_{ki})}z_{ki}^{n_{ki}}$ corresponds bijectively with the Fock state $|\text{ph}\rangle=|\{n_{ki}\}\rangle$. Let us introduce a formal state vector
\[
|z\rangle=|\{z_{ki}\}\rangle,
\]
to distinguish it from the one-dimensional variable~$z$, and define its raising to a power $|\text{ph}\rangle$ as
\[
|z\rangle^{|\text{ph}\rangle}=\prod_{k,i}z_{ki}^{n_{ki}}.
\]
Analogously, introduce a vector
\[
|\kappa\rangle=|\{\kappa_{ki}\}\rangle
\]
and a formal symbolic power
\[
|\kappa\rangle^{(|\text{ph}\rangle)}=\prod_{k,i}\kappa_{ki}^{(n_{ki})}.
\]
Then the product can be expressed as
\begin{equation}
\label{primaryF}
F_{|z\rangle}=\prod_{k,i}\sigma_{ki}(z_{ki})=\sum_{|\text{ph}\rangle}|\kappa\rangle^{(|\text{ph}\rangle)}|z\rangle^{|\text{ph}\rangle},
\end{equation}
and the correspondence between the term~$\prod\kappa_{ki}^{(n_{ki})}z_{ki}^{n_{ki}}=|\kappa\rangle^{(|\text{ph}\rangle)}|z\rangle^{|\text{ph}\rangle}$ and the state~$|\text{ph}\rangle$ becomes obvious. If we put $z_{ki}=z$ for all $k$ and~$i$, then the product $F_{|z\rangle}$ becomes a power series $F(z)$ such that the coefficient of $z^n$ gives the sum of terms $|\kappa\rangle^{(|\text{ph}\rangle)}$ over all Fock states with $n$ photons. In other words, if a quantity $A_{|\text{ph}\rangle}$ can be represented as a product $\prod\kappa_{ki}^{(n_{ki})}$,
\[
A_{|\text{ph}\rangle}=|\kappa\rangle^{(|\text{ph}\rangle)},
\]
so that
\[
F_{|z\rangle}=\sum_{|\text{ph}\rangle}A_{|\text{ph}\rangle}|z\rangle^{|\text{ph}\rangle},
\]
then
\begin{equation}
\label{FAsProductOfSigmaZ}
F(z)=\prod_{k,i}\sigma_{ki}(z)
\end{equation}
is the generating function for $\sum A_{|\text{ph}^n\rangle}$.

Since
\[
q^{\varepsilon_{|\text{ph}\rangle}}=\prod_{k=0}^\infty\prod_{i=0}^{2k+1}q^{n_{ki}k},
\]
the series
\begin{equation}
\label{finalSigmaKI}
\sigma_{ki}(z)=1+q^k z+q^{2k}z^2+\cdots=(1-q^k z)^{-1}
\end{equation}
corresponds to mode $ki$ after putting $z_{ki}=z$, and therefore all $a_n$ are determined via the generating function
\[
F(z)=\prod_{k=0}^\infty(1-q^k z)^{-g_k}.
\]
Differentiating $F(z)$ yields
\begin{equation}
\label{FDerivative}
F'(z)=\varphi(z)F(z),
\end{equation}
where
\begin{equation}
\label{phiZ}
\varphi(z)=\sum_{n=0}^\infty c_{n+1}z^n
\end{equation}
and
\begin{equation}
\label{cN}
c_n=\frac{2}{(1-q^n)^2}, \quad n\geqslant1.
\end{equation}
We have $c_n\geqslant2$. Substituting
\begin{equation}
\label{FViaAN}
F(z)=\sum_{n=0}^\infty a_n z^n
\end{equation}
together with Eq.~\eqref{phiZ} in the relation~\eqref{FDerivative}, calculating the left- and right-hand sides, and comparing terms of equal powers in $z$, we finally obtain the recursive relation for~$a_n$:
\begin{subequations}
\label{aNRecursiveRelation}
\begin{eqnarray}
a_0&=&1,
\\
a_n&=&\frac{1}{n}\sum_{m=1}^n c_m a_{n-m}, \quad n\geqslant1.
\end{eqnarray}
\end{subequations}
Thus, if we know the temperature $T$ and the energy difference between adjacent energy levels in the cavity, $\hbar\Omega$, we can determine $q$ using Eqs.~\eqref{inverseTemperature} and~\eqref{q}, then determine $c_n$ using Eq.~\eqref{cN}, and finally recursively calculate all necessary $a_n$ using Eqs.~\eqref{aNRecursiveRelation}. Note that we need to determine $a_n$ (and hence $c_n$) only for $n\leqslant n_\Sigma$; therefore, all the probabilities $\pi_n$ as well as the normalization coefficient $P$ can be found in a finite number of steps and without performing infinite summation.

Clearly, $\{\pi_n\}$ is related to~$\{\pi_{|\text{d}\rangle}\}$. Since the total excitation number is fixed, the probability of finding $n^1_{|\text{d}\rangle}$ excited dye molecules in the dye solution equals the probability of finding $n_\Sigma-n^1_{|\text{d}\rangle}$ photons in the photon gas. Multiplying the latter probability by the conditional probability of state $|\text{d}\rangle$ given $n^1_{|\text{d}\rangle}$, we have
\begin{equation}
\label{piDViaPiN}
\pi_{|\text{d}\rangle}={n_\text{d}\choose n^1_{|\text{d}\rangle}}^{-1}\pi_{n_\Sigma-n^1_{|\text{d}\rangle}}.
\end{equation}
Using Eq.~\eqref{piN} and comparing Eq.~\eqref{piDViaPiN} to Eq.~\eqref{piD1}, we see that $Q=Pr^{n_\Sigma}$. This relation can also be obtained by comparing Eqs.~\eqref{finalQ} and~\eqref{finalP} and noting that $n^\text{ph}_\text{min}+n^\text{e}_\text{max}=n_\Sigma$.

\subsection{One cavity mode}

Next consider fluctuations of the number of photons in an arbitrary mode of the microcavity, say, mode~$ki$. Denote by $|\text{ph}|ki^m\rangle$ a Fock state with $m$ mode-$ki$ photons. The probability $\pi^{ki}_m$ of finding $m$ photons in mode~$ki$ is obtained by summing $\pi_{|\text{ph}\rangle}$ over all states~$|\text{ph}|ki^m\rangle$:
\begin{equation}
\label{piKIM}
\pi^{ki}_m=\sum_{|\text{ph}|ki^m\rangle}\pi_{|\text{ph}\rangle}.
\end{equation}
Denoting by $|\text{ph}^n|ki^m\rangle$ a Fock state with a total of $n$ photons among which there are exactly $m$ mode-$ki$ photons, we may sum the probabilities~\eqref{mainPiPh} first over these states and then over all possible~$n$. The total number of photons cannot be less than the number of photons in any mode, $n\geqslant m\geqslant0$; on the other hand, it is subject to the inequality constraints~\eqref{photonNumberRange1}. We then have for $n$ that
\begin{equation}
\label{nSummationIntervalForOneMode}
n^\text{ph}_\text{min}(m)\leqslant n\leqslant n_\Sigma,
\end{equation}
where
\begin{equation}
\label{nPhMinM}
n^\text{ph}_\text{min}(m)=\max\{m,n_\Sigma-n_\text{d}\},
\end{equation}
so that $n^\text{ph}_\text{min}=n^\text{ph}_\text{min}(0)$. Consequently,
\begin{equation}
\label{piKIM1}
\pi^{ki}_m=P\!\!\!\!\sum_{n=n^\text{ph}_\text{min}(m)}^{n_\Sigma}{n_\text{d}\choose n_\Sigma-n}r^n a^{ki}_{m\,n-m},
\end{equation}
where
\begin{equation}
\label{aKIMN}
a^{ki}_{m\,n-m}=\sum_{|\text{ph}^n|ki^m\rangle}q^{\varepsilon_{|\text{ph}\rangle}},
\end{equation}
$r$ is given by Eq.~\eqref{r}, and $P$ can be calculated with Eq.~\eqref{finalP}.

Summing the right-hand side of Eq.~\eqref{aKIMN} over all possible $m$ gives the sum of $q^{\varepsilon_{|\text{ph}\rangle}}$ over all Fock states with $n$ photons, and, by the definition~\eqref{aN}, we can write
\begin{equation}
\label{aNViaAKIMN}
a_n=\sum_{m=0}^n a^{ki}_{m\,n-m}.
\end{equation}
Equation~\eqref{aNViaAKIMN} holds for all nonnegative $k$ and $n$, with $i$ in the range~\eqref{iRange}.

To find $a^{ki}_{mn}$, we use the same idea as when calculating $a_n$, but with a slight difference: When constructing the generating function
\begin{equation}
\label{FKIYZExpansion}
F^{ki}(y,z)=\sum_{m,n=0}^\infty a^{ki}_{mn}y^mz^n
\end{equation}
for the sums~\eqref{aKIMN} from the product~\eqref{primaryF}, we put $z_{ki}=y$ for mode~$ki$ and $z_{lj}=z$ for all the other modes, when $lj\neq ki$, to distinctively calculate sums for any fixed number of mode-$ki$ photons. This corresponds to replacing the series $\sigma_{ki}(z)$ for mode~$ki$ by the same series but considered as a function of~$y$, not~$z$, in Eq.~\eqref{FAsProductOfSigmaZ}. Thus,
\begin{equation}
\label{FKIYZViaFIKZ}
F^{ki}(y,z)=\frac{F^{ki}(z)}{1-q^ky},
\end{equation}
where
\begin{equation}
\label{FKIZViaFZ}
F^{ki}(z)=(1-q^kz)F(z)
\end{equation}
is the generating function giving the sums of $q^{\varepsilon_{|\text{ph}\rangle}}$ over all states with $n$ photons among which there are no mode-$ki$ photons:
\begin{equation}
\label{FIKZExpansion}
F^{ki}(z)=\sum_{n=0}^\infty a^{ki}_n z^n,
\end{equation}
where
\[
a^{ki}_n\equiv a^{ki}_{0n}=\sum_{|\text{ph}^n|ki^0\rangle}q^{\varepsilon_{|\text{ph}\rangle}}.
\]
Obviously,
\[
F^{ki}(z,z)=F(z),
\]
which reflects the relation~\eqref{aNViaAKIMN}.

Equation~\eqref{FKIYZViaFIKZ} together with the expansions~\eqref{finalSigmaKI}, \eqref{FKIYZExpansion}, and~\eqref{FIKZExpansion} yields
\[
a^{ki}_{mn}=q^{km}a^{ki}_n.
\]
Accordingly, we can rewrite the probability distribution~\eqref{piKIM1} of the number of photons in mode~$ki$ as
\begin{equation}
\label{piKIM2}
\pi^{ki}_m=Pq^{km}\!\!\!\!\sum_{n=n^\text{ph}_\text{min}(m)}^{n_\Sigma}{n_\text{d}\choose n_\Sigma-n}r^n a^{ki}_{n-m}.
\end{equation}
Using Eqs.~\eqref{FViaAN} and~\eqref{FKIZViaFZ} allows us to calculate $a^{ki}_n$:
\begin{subequations}
\label{aKINRelation}
\begin{eqnarray}
a^{ki}_0&=&1,
\\
a^{ki}_n&=&a_n-q^ka_{n-1}, \quad n\geqslant1.
\end{eqnarray}
\end{subequations}
Thus, to find the necessary $a^{ki}_n$ appearing in Eq.~\eqref{piKIM2}, we should first calculate $a_n$ for all nonnegative $n$ not exceeding $n_\Sigma$ with the recursive relation~\eqref{aNRecursiveRelation} (which we can do when calculating the normalization coefficient~$P$) and then utilize the relations~\eqref{aKINRelation}. The probability distribution $\{\pi^{ki}_m\}$ is determined by Eq.~\eqref{piKIM2} for $m$ from the range
\begin{equation}
\label{mRange}
0\leqslant m\leqslant n_\Sigma.
\end{equation}
For all the other~$m$, $\pi^{ki}_m\equiv0$.

It follows from Eqs.~\eqref{aKINRelation} that $a^{ki}_n$ in fact does not depend on~$i$, and, by Eq.~\eqref{piKIM2}, neither does~$\pi^{ki}_m$:
\begin{equation}
\label{piKIEquivalence}
\pi^{k0}_m\equiv\pi^{k1}_m\equiv\cdots\equiv\pi^{k\,2k+1}_m.
\end{equation}
Therefore, all the modes with the same energy are characterized by the same probability distribution and have the same statistical properties.

\subsection{One energy level}

Further consider fluctuations of the number of photons with a given energy. If the photon energy is $\hbar\omega_k$, we may speak of the photon number fluctuations at the $k$th energy level and define the probability $\pi^k_m$ of finding $m$ photons at this level:
\[
\pi^k_m=\sum_{|\text{ph}|k^m\rangle}\pi_{|\text{ph}\rangle},
\]
where summation is performed over all Fock states with $m$ photons at the $k$th energy level, $|\text{ph}|k^m\rangle$. By analogy with Eq.~\eqref{piKIM}, we may sum first over all Fock states with a total of $n$ photons and exactly $m$ photons with energy~$\hbar\omega_k$, $|\text{ph}^n|k^m\rangle$, and then over all possible~$n$. Clearly, the summation interval for $n$ coincides with that given by Eq.~\eqref{nSummationIntervalForOneMode}. Therefore,
\[
\pi^k_m=P\!\!\!\!\sum_{n=n^\text{ph}_\text{min}(m)}^{n_\Sigma}{n_\text{d}\choose n_\Sigma-n}r^n a^k_{m\,n-m},
\]
where
\begin{equation}
\label{aKMN}
a^k_{m\,n-m}=\sum_{|\text{ph}^n|k^m\rangle}q^{\varepsilon_{|\text{ph}\rangle}},
\end{equation}
$r$ and $n^\text{ph}_\text{min}(m)$ are defined by Eqs.~\eqref{r} and~\eqref{nPhMinM}, and $P$ can be calculated with Eq.~\eqref{finalP}.

We can write the relation analogous to Eq.~\eqref{aNViaAKIMN} because summing the right-hand side of Eq.~\eqref{aKMN} over all possible $m$ gives the sum of $q^{\varepsilon_{|\text{ph}\rangle}}$ over all Fock states with $n$ photons:
\begin{equation}
\label{aNViaAKMN}
a_n=\sum_{m=0}^n a^k_{m\,n-m}.
\end{equation}
Equation~\eqref{aNViaAKMN} holds for all nonnegative $k$ and~$n$.

To find $a^k_{mn}$, we construct the generating function
\begin{equation}
\label{FKYZExpansion}
F^k(y,z)=\sum_{m,n=0}^\infty a^k_{mn}y^mz^n
\end{equation}
by putting $z_{ki}=y$ for all $i$ from the range~\eqref{iRange} and $z_{lj}=z$ for all $l\neq k$ in $F_{|z\rangle}$ defined by Eq.~\eqref{primaryF}, which corresponds to the isolation of the states with every single number of photons with energy~$\hbar\omega_k$ among the states with a total of $n$ photons. Replacing all the series $\sigma_{ki}(z)$ corresponding to the $k$th energy level by the same series but considered as functions of~$y$ in Eq.~\eqref{FAsProductOfSigmaZ}, we get
\begin{equation}
\label{FKYZViaFKZ}
F^k(y,z)=\frac{F^k(z)}{(1-q^ky)^{g_k}},
\end{equation}
where the degeneracy $g_k$ is given by Eq.~\eqref{degeneracy} and
\begin{equation}
\label{FKZViaFZ}
F^k(z)=(1-q^kz)^{g_k}F(z)
\end{equation}
is the generating function giving the sums of $q^{\varepsilon_{|\text{ph}\rangle}}$ over all states with $n$ photons among which there are no photons with energy~$\hbar\omega_k$:
\begin{equation}
\label{FKZExpansion}
F^k(z)=\sum_{n=0}^\infty a^k_n z^n,
\end{equation}
where
\[
a^k_n\equiv a^k_{0n}=\sum_{|\text{ph}^n|k^0\rangle}q^{\varepsilon_{|\text{ph}\rangle}}.
\]
Obviously,
\[
F^k(z,z)=F(z),
\]
which reflects the relation~\eqref{aNViaAKMN}.

With the expansions~\eqref{FKYZExpansion} and~\eqref{FKZExpansion}, we obtain from Eq.~\eqref{FKYZViaFKZ}
\[
a^k_{mn}={m+g_k-1\choose m}q^{km}a^k_n.
\]
Thus, the probability distribution $\{\pi^k_m\}$ of the number of photons at the $k$th energy level has the form
\begin{equation}
\label{piKM2}
\pi^k_m=P{m+g_k-1\choose m}q^{km}\!\!\!\!
\sum_{n=n^\text{ph}_\text{min}(m)}^{n_\Sigma}
{n_\text{d}\choose n_\Sigma-n}r^n a^k_{n-m}
\end{equation}
and is determined for all $m$ from the range~\eqref{mRange}; for the other~$m$, $\pi^k_m\equiv0$. It remains to express $a^k_n$ via $a_n$ by expanding Eq.~\eqref{FKZViaFZ} and equating the coefficients of equal powers in $z$ on both sides:
\begin{equation}
\label{aKNViaAN}
a^k_n=\sum_{m=0}^{\min\{n,g_k\}}{g_k\choose m}(-1)^mq^{km}a_{n-m}.
\end{equation}
We have the situation analogous to that of one cavity mode: together with $r$ and~$q$ [Eqs.~\eqref{r} and~\eqref{q}], it is sufficient to calculate $a_n$ for nonnegative $n$ not exceeding $n_\Sigma$ using the recursive relation~\eqref{aNRecursiveRelation} to fully determine the probability distribution~$\{\pi^k_m\}$.

\subsection{Condensate fluctuations}

Using the above results, we can easily find the probability distribution describing condensate fluctuations, which are of fundamental interest in the context of Bose-Einstein condensation of light. Recall that the whole photon condensate, all the photons with energy~$\hbar\omega_0$, consists of the two polarized photon condensates, each comprising photons of one of the two polarizations. Accordingly, we distinguish between the fluctuations of the whole condensate and those of the polarized condensate.

\subsubsection{Polarized condensate}

First consider fluctuations of the polarized photon condensate, say, with polarization $p=0$. The probability $\pi^{00}_m$ of finding $m$ mode-$00$ photons follows from Eq.~\eqref{piKIM2}:
\begin{equation}
\label{polarizedCondensateProbabilityDistribution}
\pi^{00}_m=P\!\!\!\!\sum_{n=n^\text{ph}_\text{min}(m)}^{n_\Sigma}{n_\text{d}\choose n_\Sigma-n}r^n a^{00}_{n-m},
\end{equation}
where, by Eq.~\eqref{aKINRelation},
\begin{eqnarray*}
a^{00}_0&=&1,
\nonumber\\
a^{00}_n&=&a_n-a_{n-1}, \quad n\geqslant1.
\end{eqnarray*}
From the remark after Eq.~\eqref{piKIEquivalence} it follows that the second polarized photon condensate, with polarization $p=1$, has the same statistical properties as the first:
\begin{equation}
\label{identicalStatisticalPropertiesOfPolarizedCondensates}
\pi^{01}_m\equiv\pi^{00}_m.
\end{equation}

\subsubsection{Whole condensate}

Second consider fluctuations of the photon condensate as a whole. The probability $\pi^0_m$ of finding $m$ photons at the lowest energy level follows from Eq.~\eqref{piKM2}:
\begin{equation}
\label{wholeCondensateProbabilityDistribution}
\pi^0_m=P(m+1)\!\!\!\!
\sum_{n=n^\text{ph}_\text{min}(m)}^{n_\Sigma}
{n_\text{d}\choose n_\Sigma-n}r^n a^0_{n-m}.
\end{equation}
The quantities $a^0_n$ can be found directly from the relation~\eqref{aKNViaAN}, but it is reasonable to express them via $a^{00}_n$ by noting that $F^0(z)=(1-z)F^{00}(z)$:
\begin{eqnarray*}
a^0_0&=&1,
\nonumber\\
a^0_n&=&a^{00}_n-a^{00}_{n-1}, \quad n\geqslant1.
\end{eqnarray*}

\section{Universal relation}

In Sec.~VI\,D, we have obtained the probability distributions of the photon number for the polarized and whole photon condensates. From these distributions we can calculate all the moments. Let us turn our attention to the correlation characteristics of the condensates. Consider the quantum degree of second-order coherence, or the normalized zero-delay second-order correlation function \cite{Glauber1963,Glauber1963b,Kolobov1999,Loudon2000}, for the photon condensate,
\begin{equation}
\label{wholeCondensateCoherenceDegree}
g^{(2)}_0(0)=\frac{\langle n_0(n_0-1)\rangle}{\langle n_0\rangle^2},
\end{equation}
and for the polarized photon condensate,
\begin{equation}
\label{polarizedCondensateCoherenceDegree}
g^{(2)}_{00}(0)=\frac{\langle n_{00}(n_{00}-1)\rangle}{\langle n_{00}\rangle^2},
\end{equation}
where $n_0$ and $n_{00}$ are the fluctuating photon numbers corresponding to these condensates. The degrees of second-order coherence for these condensates are not independent but related to each other via the universal relation \cite{Sobyanin2013a}
\begin{equation}
\label{universalRelation}
g^{(2)}_0(0)=\frac{3}{4}\,g^{(2)}_{00}(0).
\end{equation}
The universality means that the relation~\eqref{universalRelation} holds for all numbers of dye molecules, photons, and excitations at all temperatures.

Now we prove the universal relation~\eqref{universalRelation}. We can directly check that the sum in Eq.~\eqref{wholeCondensateProbabilityDistribution} for the photon number probability distribution for the condensate,
\[
A^0_m=P\!\!\!\!\sum_{n=n^\text{ph}_\text{min}(m)}^{n_\Sigma}{n_\text{d}\choose n_\Sigma-n}r^n a^0_{n-m},
\]
is related to the photon number probabilities~\eqref{polarizedCondensateProbabilityDistribution} for the polarized condensate as follows:
\[
A^0_m=\pi^{00}_m-\pi^{00}_{m+1}.
\]
The second factorial moment for the whole condensate,
\[
\langle n_0(n_0-1)\rangle=\sum_{m=0}^{n_\Sigma}\pi^0_m m(m-1),
\]
where
\[
\pi^0_m=A^0_m(m+1),
\]
then becomes the difference between the sums of $\pi^{00}_m$ and $\pi^{00}_{m+1}$ with weight $(m-1)m(m+1)$ over all $m$ from the range~\eqref{mRange}. Increasing the variable of summation in the second sum by~$1$, we pass to summation of $\pi^{00}_{m}$ with weight $(m-2)(m-1)m$ over all $m$ from the range $1\leqslant m\leqslant n_\Sigma+1$. Observing that the latter sum remains unchanged when the summation interval is changed to $0\leqslant m\leqslant n_\Sigma$ and subtracting this sum from the former, we get
\begin{equation}
\label{secondFactorialMomentRelation}
\langle n_0(n_0-1)\rangle=3\langle n_{00}(n_{00}-1)\rangle,
\end{equation}
where
\[
\langle n_{00}(n_{00}-1)\rangle=\sum_{m=0}^{n_\Sigma}\pi^{00}_m m(m-1)
\]
is the second factorial moment for the polarized condensate. Then we should use an obvious relation between the first moments,
\begin{equation}
\label{firstMomentRelation}
\langle n_0\rangle=2\langle n_{00}\rangle,
\end{equation}
which may be obtained either formally by analogy with the relation~\eqref{secondFactorialMomentRelation} or directly averaging the equality $n_0=n_{00}+n_{01}$ and using the equivalent statistical properties of the two polarized condensates [Eq.~\eqref{identicalStatisticalPropertiesOfPolarizedCondensates}]. Substituting the relations~\eqref{secondFactorialMomentRelation} and~\eqref{firstMomentRelation} in Eq.~\eqref{wholeCondensateCoherenceDegree} and comparing the latter to Eq.~\eqref{polarizedCondensateCoherenceDegree}, we arrive at the universal relation~\eqref{universalRelation}.

The universal relation indicates that the two polarized photon condensates are, in general, not statistically independent. Indeed, in the case of independent condensates, we would have another relation between the degrees of second-order coherence~\cite{Loudon2000},
\begin{equation}
\label{relationForIndependentCondensates}
g^{(2)}_0(0)=\frac{1+g^{(2)}_{00}(0)}{2}.
\end{equation}
Therefore, there exists a correlation between the polarized condensates.

Note that the relations~\eqref{universalRelation} and~\eqref{relationForIndependentCondensates} hold simultaneously only for $g^{(2)}_{00}(0)=2$ and $g^{(2)}_0(0)=3/2$. This is natural because $g^{(2)}_{00}(0)=2$ corresponds to the usual Bose-Einstein distribution for the polarized condensate. This distribution takes place if the polarized condensate is in contact with both an ideal thermostat and an ideal photon reservoir, in which case each mode is independent of the others and the two polarized condensates are also independent.

\section{Critical temperature}

The main effect of Bose-Einstein condensation of light is accumulation of photons at the ground energy level as temperature decreases. There exists a critical temperature below which such accumulation becomes significant, and we can find this temperature from the analogy with the case of massive bosons. Indeed, in the case of condensation of $N$ bosons, the study of the temperature behavior of the boson gas in the thermodynamic limit starts with separating out the condensate:
\begin{equation}
\label{bosonNumberSeparation}
N=\langle N_0\rangle+\langle N_+\rangle,
\end{equation}
where $\langle N_0\rangle$ is the mean number of condensate bosons, bosons at the lowest energy level, and $\langle N_+\rangle$ is the mean number of bosons at higher energy levels. In the case of condensation of photons, we instead write
\begin{equation}
\label{photonNumberSeparation}
\langle n\rangle=\langle n_0\rangle+\langle n_+\rangle,
\end{equation}
where $\langle n\rangle$ is the mean total number of photons, $\langle n_0\rangle$ is the mean number of condensate photons, photons with energy~$\hbar\omega_0$, and $\langle n_+\rangle$ is the mean number of photons with energy~$\hbar\omega_k$ for positive~$k$. In Eq.~\eqref{bosonNumberSeparation}, $\langle N_+\rangle$ is the sum of mean boson numbers corresponding to Bose-Einstein distributions for energies higher than the energy of the ground state. The results of Sec.~II indicate that the spectrum of a microcavity is virtually the same as that of a two-dimensional harmonic trap with the only difference that the degeneracy in the former case is twice that in the latter. If we assume Bose-Einstein distributions for all the photon modes~$ki$ with positive~$k$, then, in Eq.~\eqref{photonNumberSeparation}, $\langle n_+\rangle$ is formally twice the sum given by~$\langle N_+\rangle$ due to the additional twofold photon polarization degeneracy. Note that when deriving the critical temperature
\begin{equation}
\label{atomicCriticalTemperature}
\tilde{T}_\text{c}=\frac{\hbar\Omega\sqrt{6N}}{\pi k_\text{B}}
\end{equation}
\begin{figure}
\includegraphics{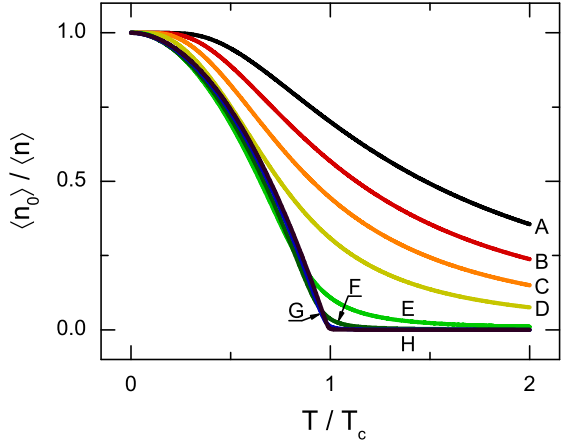}
\caption{\label{Fig1}(Color online) Condensate fraction $\langle n_0\rangle/\langle n\rangle$ against reduced temperature $T/T_\text{c}$. Parameters used: ratio of reduced partition functions: $z_1/z_0=1$; reduced detuning: $(\omega_0-\omega_\text{e})/\Omega=-10^2$; dye molecule number: $n_d=10^6$; mean photon number: $\langle n\rangle=1$ (curve~A), $2$ (B), $4$ (C), $10$ (D), $10^2$ (E), $10^3$ (F), $10^4$ (G), and $10^5$ (H).}
\end{figure}%
and temperature dependence of the condensate fraction
\begin{equation}
\label{atomicCondensateFraction}
\frac{\langle N_0\rangle}{N}=
\begin{cases}
1-\biggl(\dfrac{T}{\tilde{T}_\text{c}}\biggr)^2,&T\leqslant \tilde{T}_\text{c},\\
0,&T>\tilde{T}_\text{c},
\end{cases}
\end{equation}
in the case of the atomic BEC, we make some transformations, and all these transformations concern solely the term $\langle N_+\rangle$ \cite{BagnatoKleppner1991,HaugsetHaugerudAndersen1997,Mullin1997,WeissWilkens1997}. Analogous transformations should definitely be made in the case of the light BEC, but, fortunately, we may avoid these by formally converting Eq.~\eqref{photonNumberSeparation} to Eq.~\eqref{bosonNumberSeparation} so that $\langle n_+\rangle/2$ corresponds with $\langle N_+\rangle$. This conversion can be made using the substitutions
\[
\langle n\rangle\rightarrow2N,\quad
\langle n_+\rangle\rightarrow2\langle N_+\rangle,\quad
\langle n_0\rangle\rightarrow2\langle N_0\rangle.
\]
Thus, we need only to make the inverse substitutions
\[
N\rightarrow\frac{\langle n\rangle}{2},\quad
\langle N_0\rangle\rightarrow\frac{\langle n_0\rangle}{2}
\]
in Eqs.~\eqref{atomicCriticalTemperature} and~\eqref{atomicCondensateFraction} to obtain the critical temperature
\begin{equation}
\label{criticalTemperature}
T_\text{c} = \frac{\hbar\Omega\sqrt{3\langle n\rangle}}{\pi k_\text{B}}
\end{equation}
and temperature dependence of the condensate fraction
\begin{equation}
\label{condensateFraction}
\frac{\langle n_0\rangle}{\langle n\rangle}=
\begin{cases}
1-\biggl(\dfrac{T}{T_\text{c}}\biggr)^2,&T\leqslant T_\text{c},\\
0,&T>T_\text{c},
\end{cases}
\end{equation}
\begin{figure}
\includegraphics{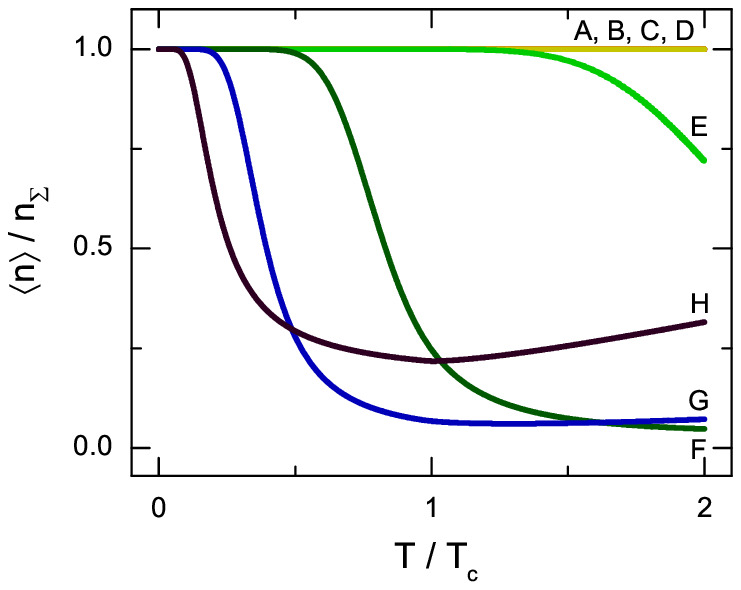}
\caption{\label{Fig2}(Color online) Photon fraction $\langle n\rangle/n_\Sigma$ against reduced temperature $T/T_\text{c}$. Parameters used are same as in Fig.~\ref{Fig1}.}
\end{figure}%
in the case of the light BEC. The dependence~\eqref{condensateFraction} reflects the true phase transition at the critical temperature~\eqref{criticalTemperature} and takes place in the thermodynamic limit, when the mean total photon number is large enough. In other words, this dependence corresponds to the macroscopic case of the light BEC. Note that Eqs.~\eqref{criticalTemperature} and~\eqref{condensateFraction} do not require invoking the above analogy with the case of the atomic BEC and can be easily derived directly from the photon number probability distributions~\eqref{piN} and~\eqref{wholeCondensateProbabilityDistribution}. In the microscopic and mesoscopic cases, when the mean total photon number is relatively small, the expression~\eqref{criticalTemperature} does not give the temperature of phase transition because this transition does not occur in a strict sense. In the latter case, the critical temperature can be interpreted as a characteristic temperature below which a significant accumulation of photons at the lowest energy level occurs. Thus, the light BEC in the microscopic and mesoscopic cases is understood in accord with the tradition of the usual atomic BEC physics \cite{Mullin1997,WeissWilkens1997}.

\section{Discussion}

\subsection{Fixed mean photon number}

It is natural to consider first a case that in some sense resembles the case of the usual atomic BEC in a two-dimensional harmonic trap, with photons playing the role of bosons of mass $m_\text{ph}$~\eqref{photonMass}. In the case of the atomic BEC, the total number of bosons is fixed. In the case of the light BEC, the total number of photons should play the role of the total number of bosons, but it is, in general, not fixed and fluctuates even at a fixed temperature. However, the mean total photon number does not fluctuate and may serve as an equivalent to the total boson number. We will thus consider first the case where the mean total photon number is fixed whereas the temperature is varied. This is especially reasonable in regard to the results of Sec.~VIII because fixing the mean total photon number implies a constant critical temperature~\eqref{criticalTemperature}, and the analogy with the atomic BEC can be readily observed.

\begin{figure}
\includegraphics{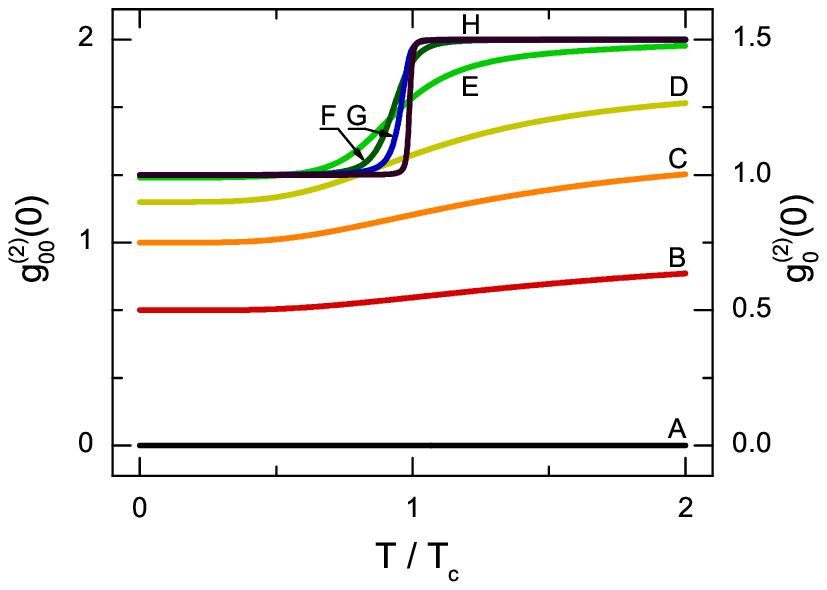}
\caption{\label{Fig3}(Color online) Degree of second-order coherence for (left axis) polarized, $g^{(2)}_{00}(0)$, and (right axis) whole, $g^{(2)}_{0}(0)$, photon condensate against reduced temperature $T/T_\text{c}$. Parameters used are same as in Fig.~\ref{Fig1}. Universal relation~\eqref{universalRelation} is seen.}
\end{figure}%

Choose the reduced detuning $(\omega_0-\omega_\text{e})/\Omega=-10^2$ and the number of dye molecules $n_\text{d}=10^6$ and study how the main characteristics of the photon gas change with temperature for the series of mean photon numbers $\langle n\rangle=1$, $2$, $4$, $10$, $10^2$, $10^3$, $10^4$, and $10^5$. This allows us to consider simultaneously the microscopic, mesoscopic, and macroscopic light BEC from a general perspective. Note that the problem of the microscopic and mesoscopic BEC is of high importance for the physics of BEC \cite{PitaevskiiStringari2003,KocharovskyEtal2006,KocharovskyKocharovsky2010}. We assume that the density of rovibrational states corresponding to the ground and first excited singlet electronic states of a dye molecule is the same: $g_0(\varepsilon)=g_1(\varepsilon)$. The corresponding reduced partition functions~\eqref{reducedPartitionFunction} are then equal to each other at all temperatures: $z_0=z_1$. It is convenient to determine the critical temperature $T_\text{c}$ with Eq.~\eqref{criticalTemperature} for each $\langle n\rangle$ and then consider the dependence of various statistical characteristics of the photon gas on the reduced temperature~$T/T_\text{c}$. The case of another reduced detuning has been considered in Ref.~\cite{Sobyanin2013a}.

Figure~\ref{Fig1} shows the temperature dependence of the condensate fraction $\langle n_0\rangle/\langle n\rangle$. We see that as temperature decreases, the photon gas undergoes BEC, and the condensate photon number becomes a macroscopic fraction of the total photon number. For relatively small $\langle n\rangle$, we observe a smooth behavior of the condensate fraction as temperature passes down through the critical value, i.e., there is no true phase transition and sharp beginning of condensation of photons. However, for $\langle n\rangle$ large enough, we observe that the light BEC sharply begins at the critical temperature and that the condensate fraction behavior is described by Eq.~\eqref{condensateFraction}. Thus, as $\langle n\rangle$ increases, the condensate fraction approaches the parabolic dependence~\eqref{condensateFraction}, so that curve H virtually coincides with the latter, and the transition to the thermodynamic limit is naturally observed. The critical temperature is the temperature of phase transition for large $\langle n\rangle$ (the macroscopic case) and the characteristic temperature at which accumulation of photons becomes significant for small~$\langle n\rangle$ (the microscopic and mesoscopic cases). Note that we speak of the phase transition and thermodynamic limit in the macroscopic case somewhat conventionally because we always deal with only a finite number of photons. At zero temperature, the condensate fraction is $1$, which indicates that all the photons are at the ground energy level for all~$\langle n\rangle$.

Figure~\ref{Fig2} shows the temperature dependence of the photon fraction $\langle n\rangle/n_\Sigma$. We see that, for curves~A--D, excitations in the optical cavity are photon excitations in the considered temperature range. For curve~E, the same situation is observed below the critical temperature, but there appears a significant fraction of excited dye molecules at high temperatures, though photons still prevail among excitations. As $\langle n\rangle$ increases further, the temperature interval in which photons are the main fraction of excitations becomes shorter, the right endpoint shifting toward zero temperature, and excited dye molecules prevail not only at high temperatures but also at temperatures below the critical temperature. Interestingly, for curves~G and~H, we observe the appearance of a temperature above which the photon fraction increases after the original decrease at low temperatures, and the minimum value of the photon fraction increases with~$\langle n\rangle$. Note that the photon fraction reflects only on average how significant photons and excited dye molecules are among excitations. The total photon number itself fluctuates, is characterized by the probability distribution~\eqref{piN}, and takes on values from the range~\eqref{photonNumberRange1}; therefore, the ratio $n/n_\Sigma$ also fluctuates and can equal~$1$ with a probability even when $\langle n\rangle/n_\Sigma\ll1$. At zero temperature, the mean total photon number coincides with the total excitation number for all natural~$\langle n\rangle$: $\langle n\rangle/n_\Sigma=1$. Therefore, the total photon number does not fluctuate at zero temperature and coincides with the total excitation number. In this connection, we may recall the Chebyshev inequality $\mathrm{P}\{\xi\geqslant x\}\leqslant\langle\xi\rangle/x$, which gives an upper estimate for the probability $\mathrm{P}\{\xi\geqslant x\}$ that a nonnegative random variable $\xi$ is no less than a positive $x$ via the ratio of the mean $\langle\xi\rangle$ to~$x$ \cite{Shiryaev1996}. Putting $\xi=n_\Sigma-n\geqslant0$, we get $\mathrm{P}\{n_\Sigma-n\geqslant x\}\leqslant(n_\Sigma-\langle n\rangle)/x$; this means that there are no fluctuations of the total photon number, $n\equiv n_\Sigma$, if $\langle n\rangle=n_\Sigma$ and that the fluctuations are small if $\langle n\rangle\approx n_\Sigma$.

\begin{figure}
\includegraphics{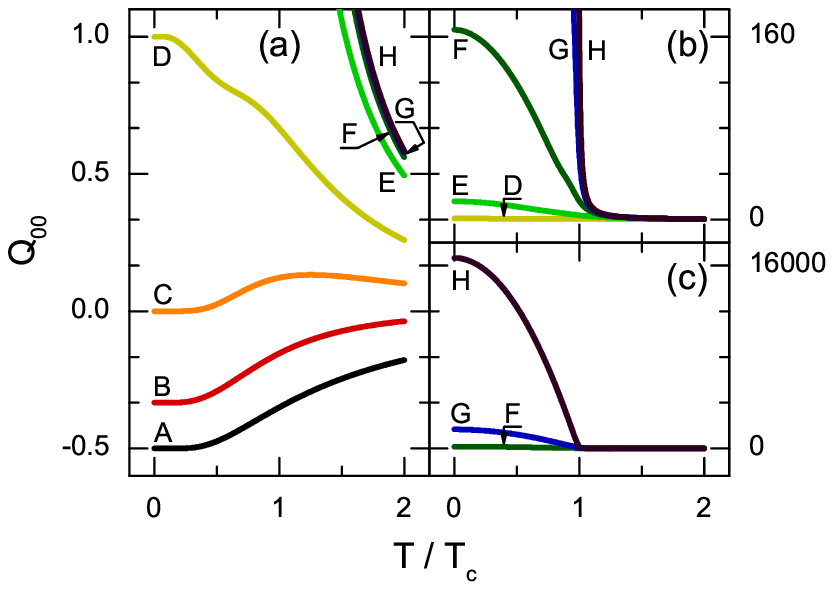}
\caption{\label{Fig4}(Color online) Mandel parameter $Q_{00}$ for polarized photon condensate against reduced temperature $T/T_\text{c}$. Parameters used are same as in Fig.~\ref{Fig1}.}
\end{figure}

Figure~\ref{Fig3} shows the temperature dependence of the degree of second-order coherence for the polarized photon condensate, $g^{(2)}_{00}(0)$, and for the whole photon condensate, $g^{(2)}_0(0)$. Each curve corresponds both to $g^{(2)}_{00}(0)$ (left axis) and to $g^{(2)}_0(0)$ (right axis), illustrating the universal relation~\eqref{universalRelation}. For any~$\langle n\rangle$, the degree of second-order coherence for the polarized and whole photon condensates increases with temperature. For large mean photon numbers, we observe Bose-Einstein statistics for the polarized photon condensate at high temperatures: $g^{(2)}_{00}(0)=2$.  As temperature decreases, $g^{(2)}_{00}(0)$ decreases but always exceeds~$1$; this means that we have super-Poissonian statistics in the whole temperature interval, even at zero temperature, where $g^{(2)}_{00}(0)=4/3$. For the whole photon condensate, we observe at first glance Poissonian but in fact sub-Poissonian statistics at low temperatures (see below), $g^{(2)}_0(0)\approx1$ but always $g^{(2)}_0(0)<1$, and super-Poissonian statistics at high temperatures, $g^{(2)}_0(0)=3/2$. The transition from the low-temperature statistics to the high-temperature statistics for the condensates becomes sharper with increasing~$\langle n\rangle$, thereby illustrating the transition to the thermodynamic limit. Note that, at high temperatures, not only the universal relation~\eqref{universalRelation} but also the relation~\eqref{relationForIndependentCondensates} holds. This means that the two polarized condensates are statistically independent at high temperatures and large mean photon numbers.

Consider the problem of sub-Poissonian statistics in more detail. This problem is important inasmuch as sub-Poissonian statistics is indicative of nonclassical states of light. Together with the degree of second-order coherence, which is less than $1$ for sub-Poissonian statistics, it is convenient to consider the Mandel parameter~\cite{Mandel1979} when investigating nonclassical states of light. For the polarized condensate it has the form
\begin{equation}
\label{Q00}
Q_{00}=\frac{\langle(\Delta n_{00})^2\rangle-\langle n_{00}\rangle}{\langle n_{00}\rangle}=(g^{(2)}_{00}(0)-1)\langle n_{00}\rangle.
\end{equation}
Analogously, for the whole condensate the Mandel parameter is
\begin{equation}
\label{Q0}
Q_0=\frac{\langle(\Delta n_0)^2\rangle-\langle n_0\rangle}{\langle n_0\rangle}=(g^{(2)}_0(0)-1)\langle n_0\rangle.
\end{equation}
In Eqs.~\eqref{Q00} and~\eqref{Q0}, $\Delta n_{00}=n_{00}-\langle n_{00}\rangle$ and $\Delta n_0=n_0-\langle n_0\rangle$ are the deviations from the mean and $\langle(\Delta n_{00})^2\rangle=\langle n_{00}^2\rangle-\langle n_{00}\rangle^2$ and $\langle(\Delta n_0)^2\rangle=\langle n_0^2\rangle-\langle n_0\rangle^2$ are the variances of the photon number for the polarized and whole condensates, respectively. The Mandel parameter is negative when the statistics is sub-Poissonian. The minimum value of the Mandel parameter is~$-1$ and is realized for a Fock state, in which the photon number is definite.

\begin{figure}
\includegraphics{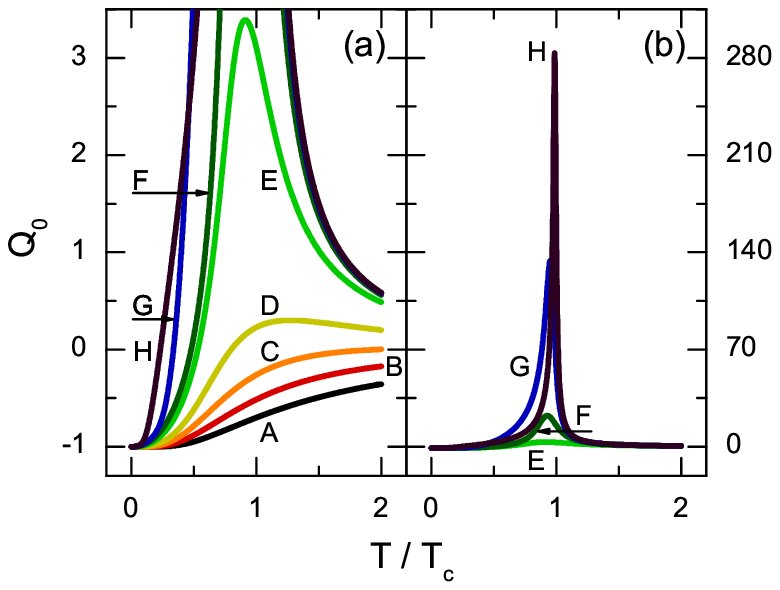}
\caption{\label{Fig5}(Color online) Mandel parameter $Q_0$ for whole photon condensate against reduced temperature $T/T_\text{c}$. Parameters used are same as in Fig.~\ref{Fig1}.}
\end{figure}

Figures~\ref{Fig4} and~\ref{Fig5} show, respectively, the temperature dependence of $Q_{00}$ and~$Q_0$. When the mean photon number is small, we observe sub-Poissonian statistics both for the polarized photon condensate ($g^{(2)}_{00}(0)<1$, $Q_{00}<0$) and for the whole photon condensate ($g^{(2)}_0(0)<1$, $Q_0<0$). This statistics takes place not only at low temperatures but also at temperatures higher than the critical temperature [see curves A and B in Figs.~\ref{Fig3}, \ref{Fig4}(a), and~\ref{Fig5}(a)]. As the mean photon number increases, there appear the states for which the statistics of the photon condensate is sub-Poissonian at some temperature whereas the statistics of the polarized photon condensate is Poissonian ($g^{(2)}_{00}(0)=1$, $Q_{00}=0$) or super-Poissonian ($g^{(2)}_{00}(0)>1$, $Q_{00}>0$); curves~C and~D in Figs.~\ref{Fig3}, \ref{Fig4}(a), and~\ref{Fig5}(a) at low temperatures illustrate these two cases. The transition from sub-Poissonian to super-Poissonian statistics for the polarized condensate at low temperatures occurs as the mean photon number passes through the value $\langle n\rangle=4$. The analogous transition also occurs at high temperatures. Note that sub-Poissonian statistics at low temperatures can disappear at higher temperatures even if $\langle n\rangle$ remains unchanged; curve~D in Figs.~\ref{Fig3} and~\ref{Fig5}(a) illustrates the transition from sub-Poissonian to super-Poissonian statistics for the whole condensate as temperature increases.

\begin{figure}
\includegraphics{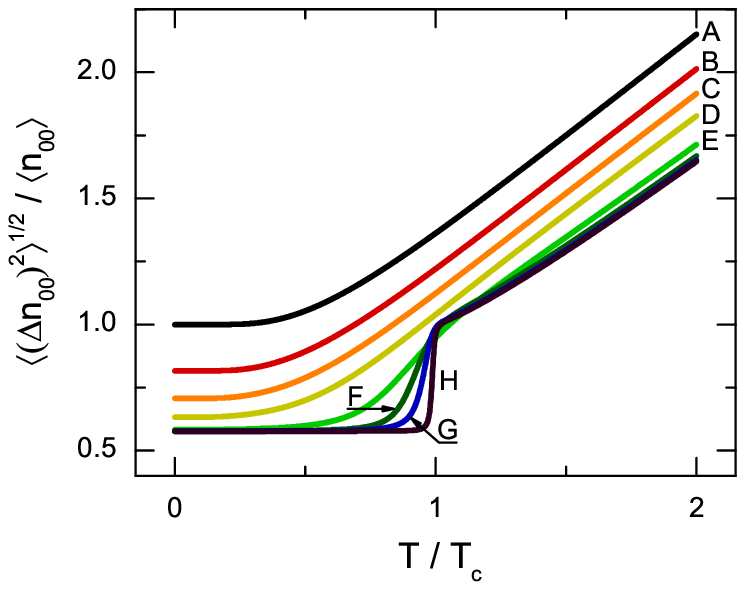}
\caption{\label{Fig6}(Color online) Reduced standard deviation $\sqrt{\langle(\Delta n_{00})^2\rangle}/\langle n_{00}\rangle$ for polarized photon condensate against reduced temperature $T/T_\text{c}$. Parameters used are same as in Fig.~\ref{Fig1}.}
\end{figure}

The statistics of the whole condensate, sub-Poissonian at small mean photon numbers, can behave in analogy to that of the polarized condensate and become Poissonian ($g^{(2)}_0(0)=1$, $Q_0=0$) and then super-Poissonian ($g^{(2)}_0(0)>1$, $Q_0>0$) at high temperatures as $\langle n\rangle$ increases [see curves A--D near $T/T_\text{c}=2$ in Figs.~\ref{Fig3} and~\ref{Fig5}(a)]. At zero temperature, however, the whole condensate is always, irrespective of~$\langle n\rangle$, in a nonclassical state with sub-Poissonian statistics: $g^{(2)}_0(0)<1$, though $g^{(2)}_0(0)$ increases with~$\langle n\rangle$ and tends to~$1$. It seems (see curves F--H in Fig.~\ref{Fig3}) that the statistics eventually becomes Poissonian, with $g^{(2)}_0(0)=1$. Nevertheless, this is not so because $Q_0=-1\neq0$, as we conclude from Fig.~\ref{Fig5}(a). Therefore, the photon condensate as a whole is in a Fock state at zero temperature for any mean total photon number.

The fact that the Mandel parameter $Q_0$ for the photon condensate tends to $-1$ as temperature vanishes naturally reflects the light BEC: All the photons at zero temperature are at the lowest energy level and have energy~$\hbar\omega_0$. The condensate photon number does not fluctuate and equals the total photon number, which coincides with the total excitation number: $n_0=n=n_\Sigma$. Indeed, it is clear from Figs.~\ref{Fig1} and~\ref{Fig2} that $\langle n_0\rangle=n_\Sigma$. Since the condensate photon number cannot exceed the total photon number, which, as we saw when discussing Fig.~\ref{Fig2}, equals the total excitation number, we have $n_0\leqslant n_\Sigma$; therefore, $n_0\equiv n_\Sigma$. Thus, the photon statistics for the whole condensate is always sub-Poissonian, not Poissonian, at zero temperature, with $Q_0=-1$ and $g^{(2)}_0(0)=1-n_\Sigma^{-1}<1$.

The polarized photon condensate, however, fluctuates at zero temperature, with $Q_{00}\geqslant-1/2>-1$ in Fig.~\ref{Fig4}(a). These fluctuations can be readily understood as follows: At zero temperature, the whole photon gas represents the condensate and comprises $n_\Sigma$ ground-mode photons, which are distributed between the two polarization states in $n_\Sigma+1$ ways with equal probability. We then have an equiprobable distribution for the polarized condensate,
\begin{equation}
\label{equiprobablePolarizedCondensateDistribution}
\pi^{00}_n=\frac{1}{n_\Sigma+1},\quad 0\leqslant n\leqslant n_\Sigma,\text{ }T=0.
\end{equation}
We immediately obtain from Eq.~\eqref{equiprobablePolarizedCondensateDistribution} the degree of second-order coherence $g^{(2)}_{00}(0)=(4/3)(1-n_\Sigma^{-1})$, evidently consistent with the universal relation~\eqref{universalRelation}, and Mandel parameter $Q_{00}=(n_\Sigma-4)/6$.

\begin{figure}
\includegraphics{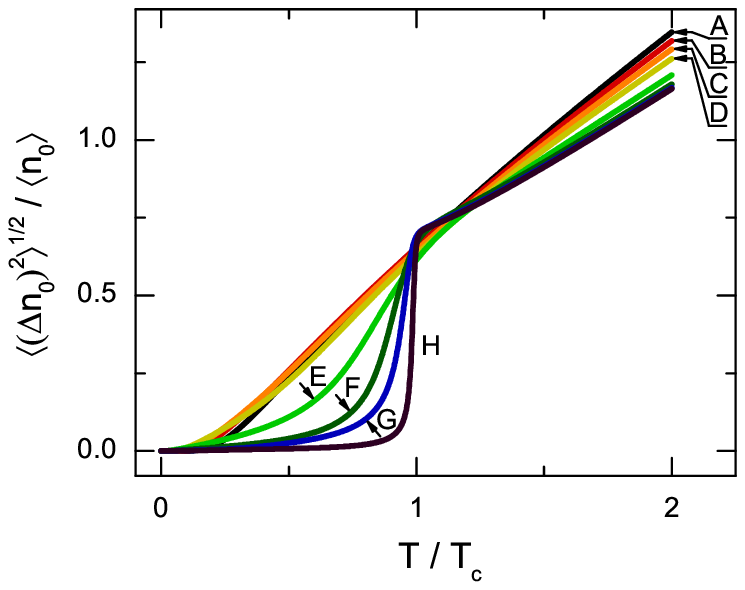}
\caption{\label{Fig7}(Color online) Reduced standard deviation $\sqrt{\langle(\Delta n_0)^2\rangle}/\langle n_0\rangle$ for whole photon condensate against reduced temperature $T/T_\text{c}$. Parameters used are same as in Fig.~\ref{Fig1}.}
\end{figure}

Note that, strictly speaking, the fluctuations of the system at absolute zero are impossible: If we put $n_{00}$ photons of one polarization and $n_{01}$ photons of the other polarization into the $\text{TEM}_{00}$ mode, with $n_{00}+n_{01}=n_\Sigma$, then these numbers will not change with time. This fact is obvious from the relation~\eqref{KSLawViaSigma}: we formally have $\sigma_{10}(\omega_0)/\sigma_{01}(\omega_0)=\infty$, which means that $\sigma_{01}(\omega_0)\equiv0$; i.e., a ground-state dye molecule cannot absorb a photon, and therefore none of the photons can change its polarization. However, the photon numbers will change with time at any infinitesimal but positive temperature, and we can consider the fluctuations of the system at zero temperature as temporal fluctuations at vanishing temperature $T\rightarrow+0$. In the ideal case of absolute zero, $T\equiv0$, we can consider the system fluctuations only as fluctuations over an ensemble of systems, each gradually cooled down to absolute zero.

For large photon numbers, $Q_{00}$ is relatively small above the critical temperature, but sharply increases as temperature passes down through the critical value, thereby indicating the beginning of BEC [see Figs.~\ref{Fig4}(b) and~\ref{Fig4}(c)]. In turn, $Q_0$ is small both at high and at low temperatures, but has a sharp peak, a cusp, near the critical temperature, in the region where the photon statistics changes [see Fig.~\ref{Fig5}(b)].

\begin{figure}
\includegraphics{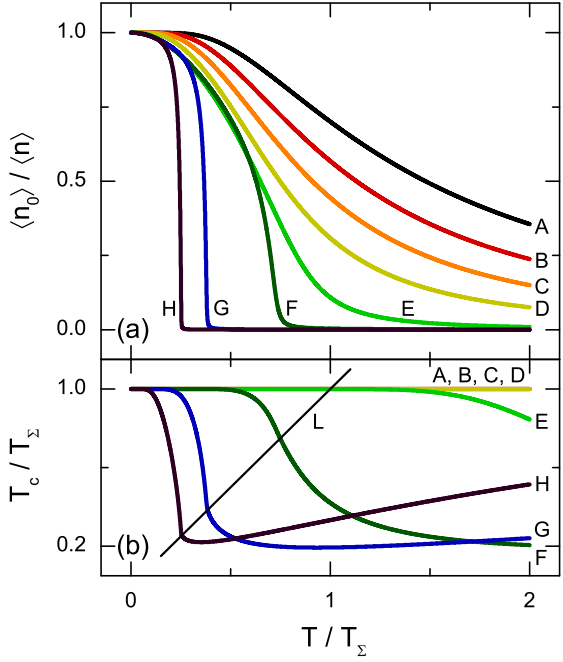}
\caption{\label{Fig8}(Color online) (a) Condensate fraction $\langle n_0\rangle/\langle n\rangle$ and (b) reduced critical temperature $T_\text{c}/T_\Sigma$ against reduced temperature $T/T_\Sigma$. Parameters used: ratio of reduced partition functions: $z_1/z_0=1$; reduced detuning: $(\omega_0-\omega_\text{e})/\Omega=-10^2$; dye molecule number: $n_d=10^6$; total excitation number: $n_\Sigma=1$ (curve~A), $2$ (B), $4$ (C), $10$ (D), $10^2$ (E), $10^3$ (F), $10^4$ (G), and $10^5$ (H). Line~L corresponds to $T=T_\text{c}$.}
\end{figure}

The difference between the fluctuating behavior of the polarized and whole condensates is also demonstrated in Figs.~\ref{Fig6} and~\ref{Fig7}, which show, respectively, the reduced standard deviation of the photon number for the polarized condensate, $\sqrt{\langle(\Delta n_{00})^2\rangle}/\langle n_{00}\rangle$, and whole condensate, $\sqrt{\langle(\Delta n_0)^2\rangle}/\langle n_0\rangle$. At zero temperature, the reduced standard deviation of the condensate photon number is zero and that of the polarized condensate photon number is positive. By Fig.~\ref{Fig1}, this implies the analogous relations for the variances: $\langle(\Delta n_0)^2\rangle=0$ and~$\langle(\Delta n_{00})^2\rangle>0$ at zero temperature, as we have already observed from the low-temperature behavior of the Mandel parameter. In the microscopic case, the reduced standard deviations both for the polarized and for the whole condensates behave smoothly as temperature passes through the critical value (see curves A--D in Figs.~\ref{Fig6} and~~\ref{Fig7}). In the mesoscopic case, there appears a slight bend near the critical temperature for both the deviations, and it sharpens and becomes a kink and eventually a jump in the macroscopic case, which reflects the transition to the thermodynamic limit (see curves E--H).

\begin{figure}
\includegraphics{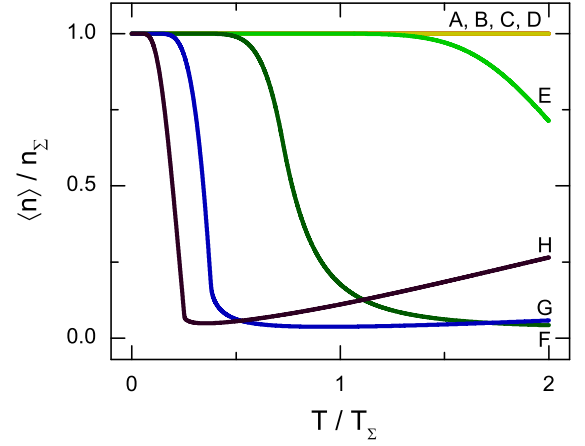}
\caption{\label{Fig9}(Color online) Photon fraction $\langle n\rangle/n_\Sigma$ against reduced temperature $T/T_\Sigma$. Parameters used are same as in Fig.~\ref{Fig8}.}
\end{figure}

The reduced standard deviation for the polarized condensate at low temperatures decreases with increasing~$\langle n\rangle$ and eventually tends to $1/\sqrt{3}\approx0.577$; it follows from Eq.~\eqref{equiprobablePolarizedCondensateDistribution} that $\sqrt{\langle(\Delta n_{00})^2\rangle}/\langle n_{00}\rangle=\sqrt{(n_\Sigma+2)/3n_\Sigma}$ at zero temperature. Simultaneously, it becomes almost constant at low temperatures, and the temperature interval in which this is so becomes larger, so that the right endpoint shifts towards the critical temperature as $\langle n\rangle$ increases. Thus, a sharp jump in $\sqrt{\langle(\Delta n_{00})^2\rangle}/\langle n_{00}\rangle$ from $1/\sqrt{3}$ to $1$ forms at the critical temperature in the macroscopic case (see curve~H in Fig.~\ref{Fig6}). Though the grand canonical approximation is never used in this paper, the latter value can be formally obtained from considering the ordinary Bose-Einstein distribution for the polarized condensate above the critical temperature, writing the corresponding reduced standard deviation $\sqrt{\langle(\Delta n_{00})^2\rangle}/\langle n_{00}\rangle=\sqrt{1+\langle n_{00}\rangle^{-1}}$, and equating the chemical potential measured from the ground-mode energy to zero at the critical temperature. This formally gives the wrong $\langle n_{00}\rangle=\infty$ but the right $\sqrt{\langle(\Delta n_{00})^2\rangle}/\langle n_{00}\rangle=1$.

The behavior of the reduced standard deviation for the whole condensate is largely similar to that for the polarized condensate. As $\langle n\rangle$ increases, $\sqrt{\langle(\Delta n_0)^2\rangle}/\langle n_0\rangle$ becomes close to zero in the temperature interval widening toward the critical temperature. Thus, we observe a jump in $\sqrt{\langle(\Delta n_0)^2\rangle}/\langle n_0\rangle$ from~$0$ to $1/\sqrt{2}\approx0.707$ near the critical temperature in the macroscopic case (see curve~H in Fig.~\ref{Fig7}). The latter value can be formally obtained by considering the two polarized photon condensates statistically independent above the critical temperature and writing $\langle(\Delta n_0)^2\rangle=2\langle(\Delta n_{00})^2\rangle$. Using Eq.~\eqref{firstMomentRelation}, we then have
\begin{equation}
\label{RMSRelation}
\frac{\sqrt{\langle(\Delta n_0)^2\rangle}}{\langle n_0\rangle}=\frac{1}{\sqrt{2}}\frac{\sqrt{\langle(\Delta n_{00})^2\rangle}}{\langle n_{00}\rangle}.
\end{equation}
For this reason, the parts of curves F--H to the right of the kink in Fig.~\ref{Fig6} are similar to the corresponding parts of curves F--H in Fig.~\ref{Fig7}.

\begin{figure}
\includegraphics{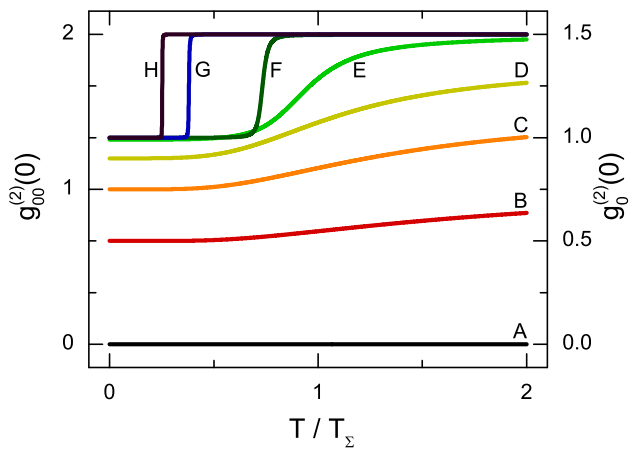}
\caption{\label{Fig10}(Color online) Degree of second-order coherence for (left axis) polarized, $g^{(2)}_{00}(0)$, and (right axis) whole, $g^{(2)}_{0}(0)$, photon condensate against reduced temperature $T/T_\Sigma$. Parameters used are same as in Fig.~\ref{Fig8}. Universal relation~\eqref{universalRelation} is seen.}
\end{figure}

\subsection{Fixed excitation number}

Now consider the case where the total number of excitations is fixed whereas the temperature is varied. We choose the same reduced detuning $(\omega_0-\omega_\text{e})/\Omega=-10^2$ and the number of dye molecules $n_\text{d}=10^6$ and study how the main characteristics of the photon gas change with temperature for the series of excitation numbers $n_\Sigma=1$, $2$, $4$, $10$, $10^2$, $10^3$, $10^4$, and $10^5$. The assumption about the density of rovibrational states is the same as earlier, so that $z_0=z_1$. The critical temperature $T_\text{c}$ is a function of the mean photon number~$\langle n\rangle$, which is not constant now but depends on temperature; this means that $T_\text{c}=T_\text{c}(T)$. We therefore normalize the temperature not by the critical temperature but by a temperature $T_\Sigma$ given by Eq.~\eqref{criticalTemperature} in which $\langle n\rangle$ is replaced by $n_\Sigma$:
\[
T_\Sigma=\frac{\hbar\Omega\sqrt{3n_\Sigma}}{\pi k_\text{B}}.
\]
The so defined $T_\Sigma$ does not depend on~$T$. We determine the temperature $T_\Sigma$ for each $n_\Sigma$ and then consider the dependence of various statistical characteristics of the photon gas on the reduced temperature~$T/T_\Sigma$.

Figure~\ref{Fig8}(a) shows the temperature dependence of the condensate fraction $\langle n_0\rangle/\langle n\rangle$. As temperature decreases, the photon gas undergoes BEC, and the condensate photon number becomes a macroscopic fraction of the total photon number. For relatively small $n_\Sigma$ (see curves A--D), the condensate fraction behaves analogously to that in the case of fixed~$\langle n\rangle$. For large~$n_\Sigma$, a crucial difference comes to the fore: as temperature decreases, Bose-Einstein condensation of light starts in a far sharper way than in the case of fixed~$\langle n\rangle$, not in a parabolic way [cf. curves F--H in Figs.~\ref{Fig1} and~\ref{Fig8}(a)]. This fact can be explained by the aforementioned dependence of the critical temperature on temperature. Indeed, $T_\Sigma=T_\text{c}$ at zero temperature. As temperature increases, absorption of photons by dye molecules begins, $\langle n\rangle$ becomes less than $n_\Sigma$, and $T_\text{c}$ decreases. Accordingly, $T/T_\text{c}$ increases nonlinearly with temperature and larger than in the case of constant~$T_\text{c}$. Figure~\ref{Fig8}(b) shows the temperature dependence of the reduced critical temperature $T_\text{c}/T_\Sigma$. Comparing Figs.~\ref{Fig8}(a) and~\ref{Fig8}(b), we see that a sharp condensation is indeed related to a sharp decrease in~$T_\text{c}$. The condensation temperature $T_\text{BEC}$ is determined from
\begin{equation}
\label{TBEC}
T_\text{BEC}=T_\text{c}(T_\text{BEC}).
\end{equation}
This condition is also suitable for the case of fixed~$\langle n\rangle$ considered above, where we have $T_\text{BEC}=T_\text{c}$ because $T_\text{c}$ is then independent of~$T$. The temperature $T_\text{BEC}$ at which the light BEC sharply starts in Fig.~\ref{Fig8}(a) corresponds to the temperature given by the intersection of the curve~$T_\text{c}/T_\Sigma$ with the line $T_\text{c}/T_\Sigma=T/T_\Sigma$, line L, in Fig.~\ref{Fig8}(b) (see curves F--H). Note that as temperature increases further, the critical temperature eventually starts increasing for curves~G and~H in Fig.~\ref{Fig8}(b).

Figure~\ref{Fig9} shows the temperature dependence of the photon fraction $\langle n\rangle/n_\Sigma$. This figure resembles Fig.~\ref{Fig8}(b) because it is the square of the latter, $\langle n\rangle/n_\Sigma=(T_\text{c}/T_\Sigma)^2$. We see that the mean photon number virtually coincides with the total excitation number for relatively small~$n_\Sigma$ (curves~A--D), and the difference arises as $n_\Sigma$ increases further. The photon fraction, as well as the critical temperature, starts increasing for curves~G and~H at high temperatures.

\begin{figure}
\includegraphics{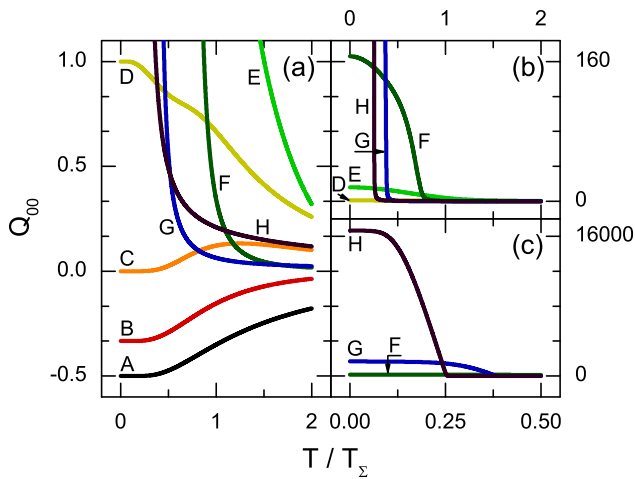}
\caption{\label{Fig11}(Color online) Mandel parameter $Q_{00}$ for polarized photon condensate against reduced temperature $T/T_\Sigma$. Parameters used are same as in Fig.~\ref{Fig8}.}
\end{figure}

\begin{figure}
\includegraphics{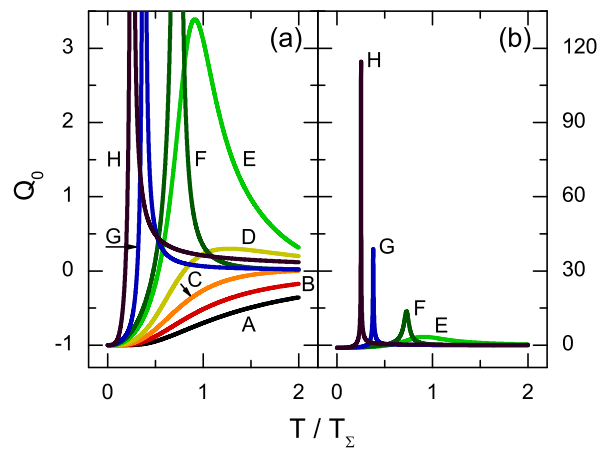}
\caption{\label{Fig12}(Color online) Mandel parameter $Q_0$ for whole photon condensate against reduced temperature $T/T_\Sigma$. Parameters used are same as in Fig.~\ref{Fig8}.}
\end{figure}

\begin{figure}
\includegraphics{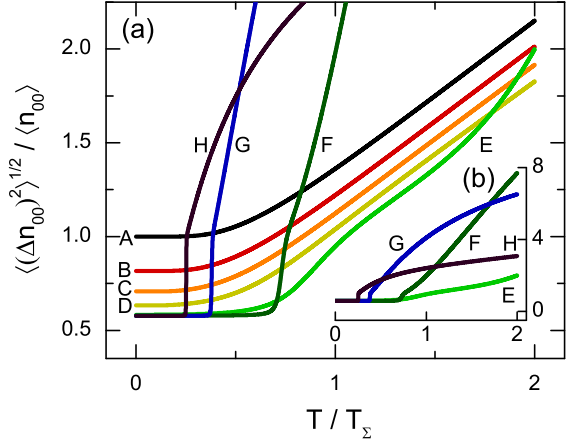}
\caption{\label{Fig13}(Color online) Reduced standard deviation $\sqrt{\langle(\Delta n_{00})^2\rangle}/\langle n_{00}\rangle$ for polarized photon condensate against reduced temperature $T/T_\Sigma$. Parameters used are same as in Fig.~\ref{Fig8}.}
\end{figure}

\begin{figure}
\includegraphics{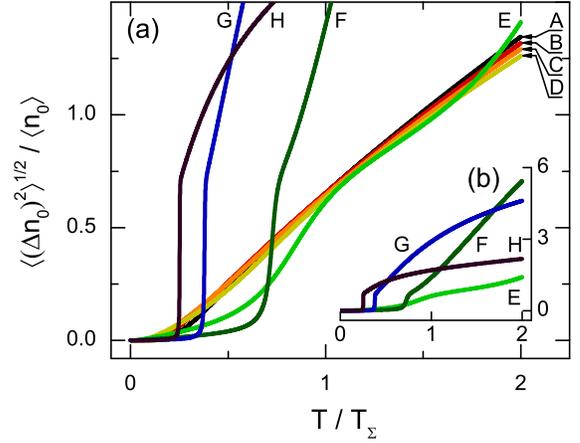}
\caption{\label{Fig14}(Color online) Reduced standard deviation $\sqrt{\langle(\Delta n_0)^2\rangle}/\langle n_0\rangle$ for whole photon condensate against reduced temperature $T/T_\Sigma$. Parameters used are same as in Fig.~\ref{Fig8}.}
\end{figure}

The temperature dependence of the degrees of second-order coherence $g^{(2)}_{00}(0)$ and  $g^{(2)}_0(0)$ is shown in Fig.~\ref{Fig10}, Mandel parameters $Q_{00}$ and~$Q_0$ in Figs.~\ref{Fig11} and~\ref{Fig12}, and reduced standard deviations $\sqrt{\langle(\Delta n_{00})^2\rangle}/\langle n_{00}\rangle$ and $\sqrt{\langle(\Delta n_0)^2\rangle}/\langle n_0\rangle$ in Figs.~\ref{Fig13} and~\ref{Fig14}. Comparing curves A--D in Figs.~\ref{Fig8}(a), \ref{Fig10}, \ref{Fig11}(a), \ref{Fig12}(a), \ref{Fig13}(a), and~\ref{Fig14}(a), respectively, with those in Figs.~\ref{Fig1}, \ref{Fig3}, \ref{Fig4}(a), \ref{Fig5}(a), \ref{Fig6}, and~\ref{Fig7}, we conclude that we cannot distinguish between the cases of fixed~$\langle n\rangle$ and fixed~$n_\Sigma$ for curves~A--D in the considered temperature interval: all the statistical characteristics are identical. The reason is the aforementioned equality $\langle n\rangle\approx n_\Sigma$ [cf. Figs.~\ref{Fig2} and~\ref{Fig9}], when virtually all excitations are photons.

In the macroscopic case, we observe the appearance of characteristics analogous to those discussed above in the case of a fixed mean photon number: In the critical region, we have a sharp jump in the degrees of second-order coherence $g^{(2)}_{00}(0)$ and $g^{(2)}_0(0)$---respectively, from $4/3$ to $2$ and from $1$ to $3/2$ in Fig.~\ref{Fig10}; we also have a sharp jump in the reduced standard deviations $\sqrt{\langle(\Delta n_{00})^2\rangle}/\langle n_{00}\rangle$ and $\sqrt{\langle(\Delta n_0)^2\rangle}/\langle n_0\rangle$---respectively, from $1/\sqrt{3}$ to $1$ and from $0$ to $1/\sqrt{2}$ in Figs.~\ref{Fig13}(a) and~\ref{Fig14}(a), and we then observe a kink. The high-temperature parts of curves F--H in Fig.~\ref{Fig13}(b) are similar to those in Fig.~\ref{Fig14}(b) because of the equality~\eqref{RMSRelation}, which reflects the independence of the two polarized photon condensates to the right of the kink. Figures~\ref{Fig11}(b) and~\ref{Fig11}(c) show that the Mandel parameter $Q_{00}$ sharply increases as temperature passes down through the critical value, and the increase occurs not in a parabolic way due to an additional dependence of $T_\text{c}$ on~$T$ [cf. Figs.~\ref{Fig4}(b) and~\ref{Fig4}(c)]. Finally, a sharp peak, a cusp, of the Mandel parameter $Q_0$ in Fig.~\ref{Fig12}(b) appears in the critical region, and it grows with~$n_\Sigma$. Thus, the temperature behavior of various statistical characteristics of the photon gas in the case of a fixed excitation number qualitatively resembles that in the case of a fixed mean total photon number, but the main feature of the former case is a much faster Bose-Einstein condensation of photons with decreasing temperature than in the latter case.

\section{Conclusion}

In this paper, I have presented the theory of Bose-Einstein condensation of light in a dye-filled optical microcavity. I have taken into account the interaction of photons in all the cavity modes with dye molecules and the twofold photon polarization degeneracy with no limitations on the numbers of photons, dye molecules, and excitations in the cavity and on the temperatures considered. The theory goes beyond the grand canonical approximation and exactly describes the microscopic, mesoscopic, and macroscopic light BEC at all temperatures, including the whole critical region.

When constructing the theory, I have used the hierarchical maximum entropy principle. In addition to this principle, I have considered an alternative, master equation approach and derived the master equation directly describing the interaction between photons and dye molecules. I have shown the equivalence between the hierarchical maximum entropy principle and the master equation approach.

I have studied fluctuations of the whole system comprising the photon gas and dye solution and derived the probability distribution of the system over all quantum states. Using the latter distribution, I have considered fluctuations of the photon gas undergoing BEC and those of the dye solution and derived the probability distribution of the photon gas over all Fock states and that of the dye solution over the numbers of ground-state and excited dye molecules. From the Fock state probability distribution, I have obtained the photon number probability distributions for the whole photon gas, each cavity mode, and each energy level. I have particularly obtained the photon number distributions describing fluctuations of the polarized and whole condensates and found the universal relation between the degrees of second-order coherence for these condensates.

I have considered two different cases of the light BEC: the case of a fixed mean total photon number and the case of a fixed total excitation number. Let the light BEC be macroscopic. Then, in the former case, the condensate fraction behaves analogously to that in the case of the usual atomic BEC and shows the parabolic temperature dependence. The crucial feature in the latter case is a much more rapid onset of BEC than in the former case, so that the condensate fraction increases much more sharply with decreasing temperature, not in a parabolic way. This difference is explained by an additional temperature dependence of the critical temperature $T_\text{c}$ in the latter case. The condensation temperature~$T_\text{BEC}$, below which BEC of photons sharply starts and which gives the phase transition temperature in the thermodynamic limit, is determined from Eq.~\eqref{TBEC}. The other behavior of the photon BEC in the two cases is quite similar: in the critical region, we observe a sharp jump in the degrees of second-order coherence and a sharp jump and kink in the reduced standard deviations for the polarized and whole condensates; a sharp peak, a cusp, of the Mandel parameter for the whole condensate; and a sharp increase in the Mandel parameter for the polarized condensate below the BEC temperature.

Significantly, the theory predicts sub-Poissonian photon statistics for the polarized and whole condensates under certain conditions. The universal relation \eqref{universalRelation} shows that sub-Poissonian photon statistics for the polarized condensate implies sub-Poissonian photon statistics for the whole condensate, but, in general, not vice versa. However, in the case of small photon numbers, we have sub-Poissonian photon statistics both for the polarized and for the whole condensates, and even at temperatures above the BEC temperature. In the case of large photon numbers, we have super-Poissonian photon statistics for the polarized condensate, but, at low temperatures, photon statistics for the whole condensate is yet sub-Poissonian. Nonclassical properties of the condensates are especially pronounced for small photon numbers and at low temperatures.

The appearance of sub-Poissonian statistics is quite natural in the context of the light BEC. Indeed, the light BEC implies accumulation of photons at the lowest energy level at zero temperature. This means that all the photon gas is the condensate, and we have an unpolarized Fock state in the $\text{TEM}_{00}$ mode of the microcavity. This Fock state is certainly nonclassical and has the degree of second-order coherence less than unity. In view of the universal relation~\eqref{universalRelation}, the degree of second-order coherence for the polarized condensate is greater by a third than that for the whole condensate; nevertheless, it can also be less than unity for the photon numbers small enough.

If the reflectivity of the mirrors is not ideal, light can escape from the microcavity. The theory predicts that this light can have unusual nonclassical properties that have not been observed so far in the light BEC experiment. Nonclassical properties of the emitted light can be investigated experimentally by measuring the degree of second-order coherence \cite{Kolobov1999}. Since the degree of second-order coherence for the light emitted by the microcavity is the same as that for the light inside the microcavity \cite{BarnettEtal1998,FosterMielkeOrozco2000}, we should observe the degree of second-order coherence for the emitted light corresponding to the $\text{TEM}_{00}$ mode less than unity under certain conditions. More concretely, to clearly observe the presence of nonclassical properties of light from these measurements, we should deal with the microscopic case of the light BEC, when we have only several photons in the microcavity; in addition, we should attempt to conduct the experiment at sufficiently low temperatures, when the reduced temperature $T/T_\text{c}$ is much less than unity. In the latter case, the state of the photon gas is expected to be close to the unpolarized Fock state in the $\text{TEM}_{00}$ mode. It is also reasonable to conduct the measurements of the degree of second-order coherence not only for the photon condensate as a whole but also for the polarized condensate by isolating one polarization mode with a polarizer in order to observe the universal relation~\eqref{universalRelation}. Thus, the behavior of the degree of second-order coherence for the polarized and whole photon condensates implies that a dye-filled optical microcavity in which the light BEC takes place can be the source of nonclassical light, both polarized and unpolarized, with sub-Poissonian photon statistics and antibunching.

I propose to observe experimentally the predictions of the theory, viz., sub-Poissonian statistics and the universal relation~\eqref{universalRelation}. Note that the experiment described in Refs.~\cite{KlaersEtal2010,KlaersEtal2011} is not suitable for observation of nonclassical light generation: In this experiment, an external laser is used to pump the microcavity and eventually to increase the number of photons confined. This technique allows one to reach the inequality $T/T_\text{c}\ll1$ by increasing the critical temperature~$T_\text{c}$ as a result of increasing the photon number whereas the microcavity remains at a constant temperature~$T$, room temperature. However, the number of photons in the microcavity becomes very large, $\approx77000$, and hence it is impossible to distinguish between sub-Poissonian and Poissonian statistics of the emitted light on the basis of measurements of the degree of second-order coherence. Accordingly, a new experimental scheme that allows one to reach small photon numbers and low reduced temperatures, or probably a modification of the old scheme and the experimental conditions, should be proposed to observe nonclassical light generation. To reach a lower reduced temperature, it might be necessary, for example, to change the temperature of the microcavity, choose the microcavity with other geometric parameters that determine $\Omega$~\eqref{Omega}, the mirror spacing and curvature, and use other fluorophores that fill the microcavity.

We finally note an interesting feature of the light BEC: under some conditions, the total number of photons in the microcavity virtually coincides with the total number of excitations, at least in some temperature interval. As we have seen, this fact implies that the fluctuations of the total photon number are very small. In this case, it is natural to expect some resemblance to the mesoscopic BEC of a fixed number of atoms in a two-dimensional harmonic trap. The question arises whether it is possible to restore any results of the theory of the mesoscopic canonical-ensemble BEC \cite{Scully1999,KocharovskyEtal2000,KocharovskyKocharovskyScully2000,ScullySvidzinsky2006,SvidzinskyScully2006,KocharovskyEtal2006,KocharovskyKocharovsky2010}
using the approach of this paper. In this connection. we might consider an imaginary situation of a microcavity with a fictitious light that has no polarization degeneracy; we would then have a microcavity the energy spectrum of which completely coincides with that of a two-dimensional harmonic trap, and we might construct a theory analogous to that described here.

Thus, the results of this paper provide an incentive for new theoretical and experimental investigations.

\begin{acknowledgments}
I would like to thank Petr Ivarovich Arseev for useful discussions.
\end{acknowledgments}
\providecommand{\noopsort}[1]{}\providecommand{\singleletter}[1]{#1}%

\begin{thebibliography}{64}%
\makeatletter
\providecommand \@ifxundefined [1]{%
 \@ifx{#1\undefined}
}%
\providecommand \@ifnum [1]{%
 \ifnum #1\expandafter \@firstoftwo
 \else \expandafter \@secondoftwo
 \fi
}%
\providecommand \@ifx [1]{%
 \ifx #1\expandafter \@firstoftwo
 \else \expandafter \@secondoftwo
 \fi
}%
\providecommand \natexlab [1]{#1}%
\providecommand \enquote  [1]{``#1''}%
\providecommand \bibnamefont  [1]{#1}%
\providecommand \bibfnamefont [1]{#1}%
\providecommand \citenamefont [1]{#1}%
\providecommand \href@noop [0]{\@secondoftwo}%
\providecommand \href [0]{\begingroup \@sanitize@url \@href}%
\providecommand \@href[1]{\@@startlink{#1}\@@href}%
\providecommand \@@href[1]{\endgroup#1\@@endlink}%
\providecommand \@sanitize@url [0]{\catcode `\\12\catcode `\$12\catcode
  `\&12\catcode `\#12\catcode `\^12\catcode `\_12\catcode `\%12\relax}%
\providecommand \@@startlink[1]{}%
\providecommand \@@endlink[0]{}%
\providecommand \url  [0]{\begingroup\@sanitize@url \@url }%
\providecommand \@url [1]{\endgroup\@href {#1}{\urlprefix }}%
\providecommand \urlprefix  [0]{URL }%
\providecommand \Eprint [0]{\href }%
\providecommand \doibase [0]{http://dx.doi.org/}%
\providecommand \selectlanguage [0]{\@gobble}%
\providecommand \bibinfo  [0]{\@secondoftwo}%
\providecommand \bibfield  [0]{\@secondoftwo}%
\providecommand \translation [1]{[#1]}%
\providecommand \BibitemOpen [0]{}%
\providecommand \bibitemStop [0]{}%
\providecommand \bibitemNoStop [0]{.\EOS\space}%
\providecommand \EOS [0]{\spacefactor3000\relax}%
\providecommand \BibitemShut  [1]{\csname bibitem#1\endcsname}%
\let\auto@bib@innerbib\@empty
\bibitem [{\citenamefont {Klaers}\ \emph {et~al.}(2010)\citenamefont {Klaers},
  \citenamefont {Schmitt}, \citenamefont {Vewinger},\ and\ \citenamefont
  {Weitz}}]{KlaersEtal2010}%
  \BibitemOpen
  \bibfield  {author} {\bibinfo {author} {\bibfnamefont {J.}~\bibnamefont
  {Klaers}}, \bibinfo {author} {\bibfnamefont {J.}~\bibnamefont {Schmitt}},
  \bibinfo {author} {\bibfnamefont {F.}~\bibnamefont {Vewinger}}, \ and\
  \bibinfo {author} {\bibfnamefont {M.}~\bibnamefont {Weitz}},\ }\href@noop {}
  {\bibfield  {journal} {\bibinfo  {journal} {Nature (London)}\ }\textbf
  {\bibinfo {volume} {468}},\ \bibinfo {pages} {545} (\bibinfo {year}
  {2010})}\BibitemShut {NoStop}%
\bibitem [{\citenamefont {Weill}\ \emph {et~al.}(2005)\citenamefont {Weill},
  \citenamefont {Rosen}, \citenamefont {Gordon}, \citenamefont {Gat},\ and\
  \citenamefont {Fischer}}]{WeillEtal2005}%
  \BibitemOpen
  \bibfield  {author} {\bibinfo {author} {\bibfnamefont {R.}~\bibnamefont
  {Weill}}, \bibinfo {author} {\bibfnamefont {A.}~\bibnamefont {Rosen}},
  \bibinfo {author} {\bibfnamefont {A.}~\bibnamefont {Gordon}}, \bibinfo
  {author} {\bibfnamefont {O.}~\bibnamefont {Gat}}, \ and\ \bibinfo {author}
  {\bibfnamefont {B.}~\bibnamefont {Fischer}},\ }\href@noop {} {\bibfield
  {journal} {\bibinfo  {journal} {Phys. Rev. Lett.}\ }\textbf {\bibinfo
  {volume} {95}},\ \bibinfo {pages} {013903} (\bibinfo {year}
  {2005})}\BibitemShut {NoStop}%
\bibitem [{\citenamefont {Connaughton}\ \emph {et~al.}(2005)\citenamefont
  {Connaughton}, \citenamefont {Josserand}, \citenamefont {Picozzi},
  \citenamefont {Pomeau},\ and\ \citenamefont {Rica}}]{ConnaughtonEtal2005}%
  \BibitemOpen
  \bibfield  {author} {\bibinfo {author} {\bibfnamefont {C.}~\bibnamefont
  {Connaughton}}, \bibinfo {author} {\bibfnamefont {C.}~\bibnamefont
  {Josserand}}, \bibinfo {author} {\bibfnamefont {A.}~\bibnamefont {Picozzi}},
  \bibinfo {author} {\bibfnamefont {Y.}~\bibnamefont {Pomeau}}, \ and\ \bibinfo
  {author} {\bibfnamefont {S.}~\bibnamefont {Rica}},\ }\href@noop {} {\bibfield
   {journal} {\bibinfo  {journal} {Phys. Rev. Lett.}\ }\textbf {\bibinfo
  {volume} {95}},\ \bibinfo {pages} {263901} (\bibinfo {year}
  {2005})}\BibitemShut {NoStop}%
\bibitem [{\citenamefont {Conti}\ \emph {et~al.}(2008)\citenamefont {Conti},
  \citenamefont {Leonetti}, \citenamefont {Fratalocchi}, \citenamefont
  {Angelani},\ and\ \citenamefont {Ruocco}}]{ContiEtal2008}%
  \BibitemOpen
  \bibfield  {author} {\bibinfo {author} {\bibfnamefont {C.}~\bibnamefont
  {Conti}}, \bibinfo {author} {\bibfnamefont {M.}~\bibnamefont {Leonetti}},
  \bibinfo {author} {\bibfnamefont {A.}~\bibnamefont {Fratalocchi}}, \bibinfo
  {author} {\bibfnamefont {L.}~\bibnamefont {Angelani}}, \ and\ \bibinfo
  {author} {\bibfnamefont {G.}~\bibnamefont {Ruocco}},\ }\href@noop {}
  {\bibfield  {journal} {\bibinfo  {journal} {Phys. Rev. Lett.}\ }\textbf
  {\bibinfo {volume} {101}},\ \bibinfo {pages} {143901} (\bibinfo {year}
  {2008})}\BibitemShut {NoStop}%
\bibitem [{\citenamefont {Novoa}\ \emph {et~al.}(2010)\citenamefont {Novoa},
  \citenamefont {Michinel}, \citenamefont {Tommasini},\ and\ \citenamefont
  {Carpentier}}]{NovoaEtal2010}%
  \BibitemOpen
  \bibfield  {author} {\bibinfo {author} {\bibfnamefont {D.}~\bibnamefont
  {Novoa}}, \bibinfo {author} {\bibfnamefont {H.}~\bibnamefont {Michinel}},
  \bibinfo {author} {\bibfnamefont {D.}~\bibnamefont {Tommasini}}, \ and\
  \bibinfo {author} {\bibfnamefont {A.~V.}\ \bibnamefont {Carpentier}},\
  }\href@noop {} {\bibfield  {journal} {\bibinfo  {journal} {Phys. Rev. A}\
  }\textbf {\bibinfo {volume} {81}},\ \bibinfo {pages} {043842} (\bibinfo
  {year} {2010})}\BibitemShut {NoStop}%
\bibitem [{\citenamefont {Weill}\ \emph
  {et~al.}(2010{\natexlab{a}})\citenamefont {Weill}, \citenamefont {Fischer},\
  and\ \citenamefont {Gat}}]{WeillFischerGat2010}%
  \BibitemOpen
  \bibfield  {author} {\bibinfo {author} {\bibfnamefont {R.}~\bibnamefont
  {Weill}}, \bibinfo {author} {\bibfnamefont {B.}~\bibnamefont {Fischer}}, \
  and\ \bibinfo {author} {\bibfnamefont {O.}~\bibnamefont {Gat}},\ }\href@noop
  {} {\bibfield  {journal} {\bibinfo  {journal} {Phys. Rev. Lett.}\ }\textbf
  {\bibinfo {volume} {104}},\ \bibinfo {pages} {173901} (\bibinfo {year}
  {2010}{\natexlab{a}})}\BibitemShut {NoStop}%
\bibitem [{\citenamefont {Weill}\ \emph
  {et~al.}(2010{\natexlab{b}})\citenamefont {Weill}, \citenamefont {Levit},
  \citenamefont {Bekker}, \citenamefont {Gat},\ and\ \citenamefont
  {Fischer}}]{WeillEtal2010}%
  \BibitemOpen
  \bibfield  {author} {\bibinfo {author} {\bibfnamefont {R.}~\bibnamefont
  {Weill}}, \bibinfo {author} {\bibfnamefont {B.}~\bibnamefont {Levit}},
  \bibinfo {author} {\bibfnamefont {A.}~\bibnamefont {Bekker}}, \bibinfo
  {author} {\bibfnamefont {O.}~\bibnamefont {Gat}}, \ and\ \bibinfo {author}
  {\bibfnamefont {B.}~\bibnamefont {Fischer}},\ }\href@noop {} {\bibfield
  {journal} {\bibinfo  {journal} {Opt. Express}\ }\textbf {\bibinfo {volume}
  {18}},\ \bibinfo {pages} {16520} (\bibinfo {year}
  {2010}{\natexlab{b}})}\BibitemShut {NoStop}%
\bibitem [{\citenamefont {Schir{\'{o}}}\ \emph {et~al.}(2012)\citenamefont
  {Schir{\'{o}}}, \citenamefont {Bordyuh}, \citenamefont {{\"{O}}ztop},\ and\
  \citenamefont {T{\"{u}}reci}}]{SchiroEtal2012}%
  \BibitemOpen
  \bibfield  {author} {\bibinfo {author} {\bibfnamefont {M.}~\bibnamefont
  {Schir{\'{o}}}}, \bibinfo {author} {\bibfnamefont {M.}~\bibnamefont
  {Bordyuh}}, \bibinfo {author} {\bibfnamefont {B.}~\bibnamefont
  {{\"{O}}ztop}}, \ and\ \bibinfo {author} {\bibfnamefont {H.~E.}\ \bibnamefont
  {T{\"{u}}reci}},\ }\href@noop {} {\bibfield  {journal} {\bibinfo  {journal}
  {Phys. Rev. Lett.}\ }\textbf {\bibinfo {volume} {109}},\ \bibinfo {pages}
  {053601} (\bibinfo {year} {2012})}\BibitemShut {NoStop}%
\bibitem [{\citenamefont {Sun}\ \emph {et~al.}(2012)\citenamefont {Sun},
  \citenamefont {Jia}, \citenamefont {Barsi}, \citenamefont {Rica},
  \citenamefont {Picozzi},\ and\ \citenamefont {Fleischer}}]{SunEtal2012}%
  \BibitemOpen
  \bibfield  {author} {\bibinfo {author} {\bibfnamefont {C.}~\bibnamefont
  {Sun}}, \bibinfo {author} {\bibfnamefont {S.}~\bibnamefont {Jia}}, \bibinfo
  {author} {\bibfnamefont {C.}~\bibnamefont {Barsi}}, \bibinfo {author}
  {\bibfnamefont {S.}~\bibnamefont {Rica}}, \bibinfo {author} {\bibfnamefont
  {A.}~\bibnamefont {Picozzi}}, \ and\ \bibinfo {author} {\bibfnamefont
  {J.~W.}\ \bibnamefont {Fleischer}},\ }\href@noop {} {\bibfield  {journal}
  {\bibinfo  {journal} {Nat. Phys.}\ }\textbf {\bibinfo {volume} {8}},\
  \bibinfo {pages} {470} (\bibinfo {year} {2012})}\BibitemShut {NoStop}%
\bibitem [{\citenamefont {Klaers}\ \emph {et~al.}(2012)\citenamefont {Klaers},
  \citenamefont {Schmitt}, \citenamefont {Damm}, \citenamefont {Vewinger},\
  and\ \citenamefont {Weitz}}]{KlaersEtal2012}%
  \BibitemOpen
  \bibfield  {author} {\bibinfo {author} {\bibfnamefont {J.}~\bibnamefont
  {Klaers}}, \bibinfo {author} {\bibfnamefont {J.}~\bibnamefont {Schmitt}},
  \bibinfo {author} {\bibfnamefont {T.}~\bibnamefont {Damm}}, \bibinfo {author}
  {\bibfnamefont {F.}~\bibnamefont {Vewinger}}, \ and\ \bibinfo {author}
  {\bibfnamefont {M.}~\bibnamefont {Weitz}},\ }\href@noop {} {\bibfield
  {journal} {\bibinfo  {journal} {Phys. Rev. Lett.}\ }\textbf {\bibinfo
  {volume} {108}},\ \bibinfo {pages} {160403} (\bibinfo {year}
  {2012})}\BibitemShut {NoStop}%
\bibitem [{\citenamefont {Sob'yanin}(2012)}]{Sobyanin2012}%
  \BibitemOpen
  \bibfield  {author} {\bibinfo {author} {\bibfnamefont {D.~N.}\ \bibnamefont
  {Sob'yanin}},\ }\href@noop {} {\bibfield  {journal} {\bibinfo  {journal}
  {Phys. Rev. E}\ }\textbf {\bibinfo {volume} {85}},\ \bibinfo {pages} {061120}
  (\bibinfo {year} {2012})}\BibitemShut {NoStop}%
\bibitem [{\citenamefont {Zhang}\ \emph
  {et~al.}(2012{\natexlab{a}})\citenamefont {Zhang}, \citenamefont {Yuan},
  \citenamefont {Zhang},\ and\ \citenamefont {Cheng}}]{ZhangEtal2012a}%
  \BibitemOpen
  \bibfield  {author} {\bibinfo {author} {\bibfnamefont {J.-J.}\ \bibnamefont
  {Zhang}}, \bibinfo {author} {\bibfnamefont {J.-H.}\ \bibnamefont {Yuan}},
  \bibinfo {author} {\bibfnamefont {J.-P.}\ \bibnamefont {Zhang}}, \ and\
  \bibinfo {author} {\bibfnamefont {Z.}~\bibnamefont {Cheng}},\ }\href@noop {}
  {\bibfield  {journal} {\bibinfo  {journal} {Commun. Theor. Phys.}\ }\textbf
  {\bibinfo {volume} {58}},\ \bibinfo {pages} {155} (\bibinfo {year}
  {2012}{\natexlab{a}})}\BibitemShut {NoStop}%
\bibitem [{\citenamefont {Zhang}\ \emph
  {et~al.}(2012{\natexlab{b}})\citenamefont {Zhang}, \citenamefont {Yuan},
  \citenamefont {Zhang},\ and\ \citenamefont {Cheng}}]{ZhangEtal2012b}%
  \BibitemOpen
  \bibfield  {author} {\bibinfo {author} {\bibfnamefont {J.-J.}\ \bibnamefont
  {Zhang}}, \bibinfo {author} {\bibfnamefont {J.-H.}\ \bibnamefont {Yuan}},
  \bibinfo {author} {\bibfnamefont {J.-P.}\ \bibnamefont {Zhang}}, \ and\
  \bibinfo {author} {\bibfnamefont {Z.}~\bibnamefont {Cheng}},\ }\href@noop {}
  {\bibfield  {journal} {\bibinfo  {journal} {Physica E}\ }\textbf {\bibinfo
  {volume} {45}},\ \bibinfo {pages} {177} (\bibinfo {year}
  {2012}{\natexlab{b}})}\BibitemShut {NoStop}%
\bibitem [{\citenamefont {Sob'yanin}(2013)}]{Sobyanin2013a}%
  \BibitemOpen
  \bibfield  {author} {\bibinfo {author} {\bibfnamefont {D.~N.}\ \bibnamefont
  {Sob'yanin}},\ }\href@noop {} {\bibfield  {journal} {\bibinfo  {journal}
  {Bull. Lebedev Phys. Inst.}\ }\textbf {\bibinfo {volume} {40}},\ \bibinfo
  {pages} {91} (\bibinfo {year} {2013})}\BibitemShut {NoStop}%
\bibitem [{\citenamefont {Zhang}\ \emph {et~al.}(2013)\citenamefont {Zhang},
  \citenamefont {Yuan}, \citenamefont {Zhang},\ and\ \citenamefont
  {Cheng}}]{ZhangEtal2013}%
  \BibitemOpen
  \bibfield  {author} {\bibinfo {author} {\bibfnamefont {J.-J.}\ \bibnamefont
  {Zhang}}, \bibinfo {author} {\bibfnamefont {J.-H.}\ \bibnamefont {Yuan}},
  \bibinfo {author} {\bibfnamefont {J.-P.}\ \bibnamefont {Zhang}}, \ and\
  \bibinfo {author} {\bibfnamefont {Z.}~\bibnamefont {Cheng}},\ }\href@noop {}
  {\bibfield  {journal} {\bibinfo  {journal} {Physica B}\ }\textbf {\bibinfo
  {volume} {408}},\ \bibinfo {pages} {16} (\bibinfo {year} {2013})}\BibitemShut
  {NoStop}%
\bibitem [{\citenamefont {Snoke}\ and\ \citenamefont
  {Girvin}(2013)}]{SnokeGirvin2013}%
  \BibitemOpen
  \bibfield  {author} {\bibinfo {author} {\bibfnamefont {D.~W.}\ \bibnamefont
  {Snoke}}\ and\ \bibinfo {author} {\bibfnamefont {S.~M.}\ \bibnamefont
  {Girvin}},\ }\href@noop {} {\bibfield  {journal} {\bibinfo  {journal} {J. Low
  Temp. Phys.}\ }\textbf {\bibinfo {volume} {171}},\ \bibinfo {pages} {1}
  (\bibinfo {year} {2013})}\BibitemShut {NoStop}%
\bibitem [{\citenamefont {Kruchkov}\ and\ \citenamefont {\relax{Yu}.
  Slyusarenko}(2013)}]{KruchkovSlyusarenko2013}%
  \BibitemOpen
  \bibfield  {author} {\bibinfo {author} {\bibfnamefont {A.}~\bibnamefont
  {Kruchkov}}\ and\ \bibinfo {author} {\bibnamefont {\relax{Yu}.
  Slyusarenko}},\ }\href@noop {} {\bibfield  {journal} {\bibinfo  {journal}
  {Phys. Rev. A}\ }\textbf {\bibinfo {volume} {88}},\ \bibinfo {pages} {013615}
  (\bibinfo {year} {2013})}\BibitemShut {NoStop}%
\bibitem [{\citenamefont {Kirton}\ and\ \citenamefont
  {Keeling}()}]{KirtonKeeling2013}%
  \BibitemOpen
  \bibfield  {author} {\bibinfo {author} {\bibfnamefont {P.}~\bibnamefont
  {Kirton}}\ and\ \bibinfo {author} {\bibfnamefont {J.}~\bibnamefont
  {Keeling}},\ }\href@noop {} {}\Eprint {http://arxiv.org/abs/1303.3459}
  {arXiv:1303.3459} \BibitemShut {NoStop}%
\bibitem [{\citenamefont {\relax{de} Leeuw}\ \emph {et~al.}()\citenamefont
  {\relax{de} Leeuw}, \citenamefont {Stoof},\ and\ \citenamefont
  {Duine}}]{LeeuwStoofDuine2013}%
  \BibitemOpen
  \bibfield  {author} {\bibinfo {author} {\bibfnamefont {A.-W.}\ \bibnamefont
  {\relax{de} Leeuw}}, \bibinfo {author} {\bibfnamefont {H.~T.~C.}\
  \bibnamefont {Stoof}}, \ and\ \bibinfo {author} {\bibfnamefont {R.~A.}\
  \bibnamefont {Duine}},\ }\href@noop {} {}\Eprint
  {http://arxiv.org/abs/1306.5107} {arXiv:1306.5107} \BibitemShut {NoStop}%
\bibitem [{\citenamefont {Boyd}\ and\ \citenamefont
  {Gordon}(1961)}]{BoydGordon1961}%
  \BibitemOpen
  \bibfield  {author} {\bibinfo {author} {\bibfnamefont {G.~D.}\ \bibnamefont
  {Boyd}}\ and\ \bibinfo {author} {\bibfnamefont {J.~P.}\ \bibnamefont
  {Gordon}},\ }\href@noop {} {\bibfield  {journal} {\bibinfo  {journal} {Bell
  Syst. Tech. J.}\ }\textbf {\bibinfo {volume} {40}},\ \bibinfo {pages} {489}
  (\bibinfo {year} {1961})}\BibitemShut {NoStop}%
\bibitem [{\citenamefont {Boyd}\ and\ \citenamefont
  {Kogelnik}(1962)}]{BoydKogelnik1962}%
  \BibitemOpen
  \bibfield  {author} {\bibinfo {author} {\bibfnamefont {G.~D.}\ \bibnamefont
  {Boyd}}\ and\ \bibinfo {author} {\bibfnamefont {H.}~\bibnamefont
  {Kogelnik}},\ }\href@noop {} {\bibfield  {journal} {\bibinfo  {journal} {Bell
  Syst. Tech. J.}\ }\textbf {\bibinfo {volume} {41}},\ \bibinfo {pages} {1347}
  (\bibinfo {year} {1962})}\BibitemShut {NoStop}%
\bibitem [{\citenamefont {Vainshtein}(1963)}]{Vainshtein1963}%
  \BibitemOpen
  \bibfield  {author} {\bibinfo {author} {\bibfnamefont {L.~A.}\ \bibnamefont
  {Vainshtein}},\ }\href@noop {} {\bibfield  {journal} {\bibinfo  {journal}
  {Zh. Eksp. Teor. Fiz.}\ }\textbf {\bibinfo {volume} {45}},\ \bibinfo {pages}
  {684} (\bibinfo {year} {1963})}\BibitemShut {NoStop}%
\bibitem [{\citenamefont {Vainshtein}(1966)}]{Vainshtein1966}%
  \BibitemOpen
  \bibfield  {author} {\bibinfo {author} {\bibfnamefont {L.~A.}\ \bibnamefont
  {Vainshtein}},\ }\href@noop {} {\emph {\bibinfo {title} {Open Resonators and
  Open Waveguides}}}\ (\bibinfo  {publisher} {Sovetskoe Radio},\ \bibinfo
  {address} {Moscow},\ \bibinfo {year} {1966})\BibitemShut {NoStop}%
\bibitem [{\citenamefont {Kogelnik}\ and\ \citenamefont
  {Li}(1966{\natexlab{a}})}]{KogelnikLi1966a}%
  \BibitemOpen
  \bibfield  {author} {\bibinfo {author} {\bibfnamefont {H.}~\bibnamefont
  {Kogelnik}}\ and\ \bibinfo {author} {\bibfnamefont {T.}~\bibnamefont {Li}},\
  }\href@noop {} {\bibfield  {journal} {\bibinfo  {journal} {Proc. IEEE}\
  }\textbf {\bibinfo {volume} {54}},\ \bibinfo {pages} {1312} (\bibinfo {year}
  {1966}{\natexlab{a}})}\BibitemShut {NoStop}%
\bibitem [{\citenamefont {Kogelnik}\ and\ \citenamefont
  {Li}(1966{\natexlab{b}})}]{KogelnikLi1966b}%
  \BibitemOpen
  \bibfield  {author} {\bibinfo {author} {\bibfnamefont {H.}~\bibnamefont
  {Kogelnik}}\ and\ \bibinfo {author} {\bibfnamefont {T.}~\bibnamefont {Li}},\
  }\href@noop {} {\bibfield  {journal} {\bibinfo  {journal} {Appl. Opt.}\
  }\textbf {\bibinfo {volume} {5}},\ \bibinfo {pages} {1550} (\bibinfo {year}
  {1966}{\natexlab{b}})}\BibitemShut {NoStop}%
\bibitem [{\citenamefont {Klaers}\ \emph {et~al.}(2011)\citenamefont {Klaers},
  \citenamefont {Schmitt}, \citenamefont {Damm}, \citenamefont {Vewinger},\
  and\ \citenamefont {Weitz}}]{KlaersEtal2011}%
  \BibitemOpen
  \bibfield  {author} {\bibinfo {author} {\bibfnamefont {J.}~\bibnamefont
  {Klaers}}, \bibinfo {author} {\bibfnamefont {J.}~\bibnamefont {Schmitt}},
  \bibinfo {author} {\bibfnamefont {T.}~\bibnamefont {Damm}}, \bibinfo {author}
  {\bibfnamefont {F.}~\bibnamefont {Vewinger}}, \ and\ \bibinfo {author}
  {\bibfnamefont {M.}~\bibnamefont {Weitz}},\ }\href@noop {} {\bibfield
  {journal} {\bibinfo  {journal} {Appl. Phys. B}\ }\textbf {\bibinfo {volume}
  {105}},\ \bibinfo {pages} {17} (\bibinfo {year} {2011})}\BibitemShut
  {NoStop}%
\bibitem [{\citenamefont {Demokritov}\ \emph {et~al.}(2006)\citenamefont
  {Demokritov}, \citenamefont {Demidov}, \citenamefont {Dzyapko}, \citenamefont
  {Melkov}, \citenamefont {Serga}, \citenamefont {Hillebrands},\ and\
  \citenamefont {Slavin}}]{DemokritovEtal2006}%
  \BibitemOpen
  \bibfield  {author} {\bibinfo {author} {\bibfnamefont {S.~O.}\ \bibnamefont
  {Demokritov}}, \bibinfo {author} {\bibfnamefont {V.~E.}\ \bibnamefont
  {Demidov}}, \bibinfo {author} {\bibfnamefont {O.}~\bibnamefont {Dzyapko}},
  \bibinfo {author} {\bibfnamefont {G.~A.}\ \bibnamefont {Melkov}}, \bibinfo
  {author} {\bibfnamefont {A.~A.}\ \bibnamefont {Serga}}, \bibinfo {author}
  {\bibfnamefont {B.}~\bibnamefont {Hillebrands}}, \ and\ \bibinfo {author}
  {\bibfnamefont {A.~N.}\ \bibnamefont {Slavin}},\ }\href@noop {} {\bibfield
  {journal} {\bibinfo  {journal} {Nature (London)}\ }\textbf {\bibinfo {volume}
  {443}},\ \bibinfo {pages} {430} (\bibinfo {year} {2006})}\BibitemShut
  {NoStop}%
\bibitem [{\citenamefont {Chumak}\ \emph {et~al.}(2009)\citenamefont {Chumak},
  \citenamefont {Melkov}, \citenamefont {Demidov}, \citenamefont {Dzyapko},
  \citenamefont {Safonov},\ and\ \citenamefont {Demokritov}}]{ChumakEtal2009}%
  \BibitemOpen
  \bibfield  {author} {\bibinfo {author} {\bibfnamefont {A.~V.}\ \bibnamefont
  {Chumak}}, \bibinfo {author} {\bibfnamefont {G.~A.}\ \bibnamefont {Melkov}},
  \bibinfo {author} {\bibfnamefont {V.~E.}\ \bibnamefont {Demidov}}, \bibinfo
  {author} {\bibfnamefont {O.}~\bibnamefont {Dzyapko}}, \bibinfo {author}
  {\bibfnamefont {V.~L.}\ \bibnamefont {Safonov}}, \ and\ \bibinfo {author}
  {\bibfnamefont {S.~O.}\ \bibnamefont {Demokritov}},\ }\href@noop {}
  {\bibfield  {journal} {\bibinfo  {journal} {Phys. Rev. Lett.}\ }\textbf
  {\bibinfo {volume} {102}},\ \bibinfo {pages} {187205} (\bibinfo {year}
  {2009})}\BibitemShut {NoStop}%
\bibitem [{\citenamefont {Malomed}\ \emph {et~al.}(2010)\citenamefont
  {Malomed}, \citenamefont {Dzyapko}, \citenamefont {Demidov},\ and\
  \citenamefont {Demokritov}}]{MalomedEtal2010}%
  \BibitemOpen
  \bibfield  {author} {\bibinfo {author} {\bibfnamefont {B.~A.}\ \bibnamefont
  {Malomed}}, \bibinfo {author} {\bibfnamefont {O.}~\bibnamefont {Dzyapko}},
  \bibinfo {author} {\bibfnamefont {V.~E.}\ \bibnamefont {Demidov}}, \ and\
  \bibinfo {author} {\bibfnamefont {S.~O.}\ \bibnamefont {Demokritov}},\
  }\href@noop {} {\bibfield  {journal} {\bibinfo  {journal} {Phys. Rev. B}\
  }\textbf {\bibinfo {volume} {81}},\ \bibinfo {pages} {024418} (\bibinfo
  {year} {2010})}\BibitemShut {NoStop}%
\bibitem [{\citenamefont {Shank}(1975)}]{Shank1975}%
  \BibitemOpen
  \bibfield  {author} {\bibinfo {author} {\bibfnamefont {C.~V.}\ \bibnamefont
  {Shank}},\ }\href@noop {} {\bibfield  {journal} {\bibinfo  {journal} {Rev.
  Mod. Phys.}\ }\textbf {\bibinfo {volume} {47}},\ \bibinfo {pages} {649}
  (\bibinfo {year} {1975})}\BibitemShut {NoStop}%
\bibitem [{\citenamefont {Schaefer}\ and\ \citenamefont
  {Willis}(1976)}]{SchaeferWillis1976}%
  \BibitemOpen
  \bibfield  {author} {\bibinfo {author} {\bibfnamefont {R.~B.}\ \bibnamefont
  {Schaefer}}\ and\ \bibinfo {author} {\bibfnamefont {C.~R.}\ \bibnamefont
  {Willis}},\ }\href@noop {} {\bibfield  {journal} {\bibinfo  {journal} {Phys.
  Rev. A}\ }\textbf {\bibinfo {volume} {13}},\ \bibinfo {pages} {1874}
  (\bibinfo {year} {1976})}\BibitemShut {NoStop}%
\bibitem [{\citenamefont {Haas}\ and\ \citenamefont
  {Rotter}(1991)}]{HaasRotter1991}%
  \BibitemOpen
  \bibfield  {author} {\bibinfo {author} {\bibfnamefont {R.~A.}\ \bibnamefont
  {Haas}}\ and\ \bibinfo {author} {\bibfnamefont {M.~D.}\ \bibnamefont
  {Rotter}},\ }\href@noop {} {\bibfield  {journal} {\bibinfo  {journal} {Phys.
  Rev. A}\ }\textbf {\bibinfo {volume} {43}},\ \bibinfo {pages} {1573}
  (\bibinfo {year} {1991})}\BibitemShut {NoStop}%
\bibitem [{\citenamefont {Schaefer}(1990)}]{Schaefer1990}%
  \BibitemOpen
  \bibinfo {editor} {\bibfnamefont {F.~P.}\ \bibnamefont {Schaefer}},\ ed.,\
  \href@noop {} {\emph {\bibinfo {title} {Dye Lasers}}},\ \bibinfo {edition}
  {3rd}\ ed.,\ \bibinfo {series} {Topics in Applied Physics}, Vol.~\bibinfo
  {volume} {1}\ (\bibinfo  {publisher} {Springer},\ \bibinfo {address}
  {Berlin},\ \bibinfo {year} {1990})\BibitemShut {NoStop}%
\bibitem [{\citenamefont {Lakowicz}(2006)}]{Lakowicz2006}%
  \BibitemOpen
  \bibfield  {author} {\bibinfo {author} {\bibfnamefont {J.~R.}\ \bibnamefont
  {Lakowicz}},\ }\href@noop {} {\emph {\bibinfo {title} {Principles of
  Fluorescence Spectroscopy}}},\ \bibinfo {edition} {3rd}\ ed.\ (\bibinfo
  {publisher} {Springer},\ \bibinfo {address} {New York},\ \bibinfo {year}
  {2006})\BibitemShut {NoStop}%
\bibitem [{\citenamefont {Sob'yanin}(2011)}]{Sobyanin2011}%
  \BibitemOpen
  \bibfield  {author} {\bibinfo {author} {\bibfnamefont {D.~N.}\ \bibnamefont
  {Sob'yanin}},\ }\href@noop {} {\bibfield  {journal} {\bibinfo  {journal}
  {Phys. Rev. E}\ }\textbf {\bibinfo {volume} {84}},\ \bibinfo {pages} {051128}
  (\bibinfo {year} {2011})}\BibitemShut {NoStop}%
\bibitem [{\citenamefont {Beck}\ and\ \citenamefont
  {Cohen}(2003)}]{BeckCohen2003}%
  \BibitemOpen
  \bibfield  {author} {\bibinfo {author} {\bibfnamefont {C.}~\bibnamefont
  {Beck}}\ and\ \bibinfo {author} {\bibfnamefont {E.~G.~D.}\ \bibnamefont
  {Cohen}},\ }\href@noop {} {\bibfield  {journal} {\bibinfo  {journal} {Physica
  A}\ }\textbf {\bibinfo {volume} {322}},\ \bibinfo {pages} {267} (\bibinfo
  {year} {2003})}\BibitemShut {NoStop}%
\bibitem [{\citenamefont {Crooks}(2007)}]{Crooks2007}%
  \BibitemOpen
  \bibfield  {author} {\bibinfo {author} {\bibfnamefont {G.~E.}\ \bibnamefont
  {Crooks}},\ }\href@noop {} {\bibfield  {journal} {\bibinfo  {journal} {Phys.
  Rev. E}\ }\textbf {\bibinfo {volume} {75}},\ \bibinfo {pages} {041119}
  (\bibinfo {year} {2007})}\BibitemShut {NoStop}%
\bibitem [{\citenamefont {Abe}\ \emph {et~al.}(2007)\citenamefont {Abe},
  \citenamefont {Beck},\ and\ \citenamefont {Cohen}}]{AbeBeckCohen2007}%
  \BibitemOpen
  \bibfield  {author} {\bibinfo {author} {\bibfnamefont {S.}~\bibnamefont
  {Abe}}, \bibinfo {author} {\bibfnamefont {C.}~\bibnamefont {Beck}}, \ and\
  \bibinfo {author} {\bibfnamefont {E.~G.~D.}\ \bibnamefont {Cohen}},\
  }\href@noop {} {\bibfield  {journal} {\bibinfo  {journal} {Phys. Rev. E}\
  }\textbf {\bibinfo {volume} {76}},\ \bibinfo {pages} {031102} (\bibinfo
  {year} {2007})}\BibitemShut {NoStop}%
\bibitem [{\citenamefont {\relax{Van der Straeten}}\ and\ \citenamefont
  {Beck}(2008)}]{StraetenBeck2008}%
  \BibitemOpen
  \bibfield  {author} {\bibinfo {author} {\bibfnamefont {E.}~\bibnamefont
  {\relax{Van der Straeten}}}\ and\ \bibinfo {author} {\bibfnamefont
  {C.}~\bibnamefont {Beck}},\ }\href@noop {} {\bibfield  {journal} {\bibinfo
  {journal} {Phys. Rev. E}\ }\textbf {\bibinfo {volume} {78}},\ \bibinfo
  {pages} {051101} (\bibinfo {year} {2008})}\BibitemShut {NoStop}%
\bibitem [{\citenamefont {Abe}(2009)}]{Abe2009}%
  \BibitemOpen
  \bibfield  {author} {\bibinfo {author} {\bibfnamefont {S.}~\bibnamefont
  {Abe}},\ }\href@noop {} {\bibfield  {journal} {\bibinfo  {journal} {Cent.
  Eur. J. Phys.}\ }\textbf {\bibinfo {volume} {7}},\ \bibinfo {pages} {401}
  (\bibinfo {year} {2009})}\BibitemShut {NoStop}%
\bibitem [{\citenamefont {Abe}(2010)}]{Abe2010}%
  \BibitemOpen
  \bibfield  {author} {\bibinfo {author} {\bibfnamefont {S.}~\bibnamefont
  {Abe}},\ }\href@noop {} {\bibfield  {journal} {\bibinfo  {journal} {Phys.
  Rev. E}\ }\textbf {\bibinfo {volume} {82}},\ \bibinfo {pages} {011131}
  (\bibinfo {year} {2010})}\BibitemShut {NoStop}%
\bibitem [{\citenamefont {Scully}(1999)}]{Scully1999}%
  \BibitemOpen
  \bibfield  {author} {\bibinfo {author} {\bibfnamefont {M.~O.}\ \bibnamefont
  {Scully}},\ }\href@noop {} {\bibfield  {journal} {\bibinfo  {journal} {Phys.
  Rev. Lett.}\ }\textbf {\bibinfo {volume} {82}},\ \bibinfo {pages} {3927}
  (\bibinfo {year} {1999})}\BibitemShut {NoStop}%
\bibitem [{\citenamefont {Kocharovsky}\ \emph
  {et~al.}(2000{\natexlab{a}})\citenamefont {Kocharovsky}, \citenamefont
  {Scully}, \citenamefont {Zhu},\ and\ \citenamefont
  {Zubairy}}]{KocharovskyEtal2000}%
  \BibitemOpen
  \bibfield  {author} {\bibinfo {author} {\bibfnamefont {V.~V.}\ \bibnamefont
  {Kocharovsky}}, \bibinfo {author} {\bibfnamefont {M.~O.}\ \bibnamefont
  {Scully}}, \bibinfo {author} {\bibfnamefont {S.-Y.}\ \bibnamefont {Zhu}}, \
  and\ \bibinfo {author} {\bibfnamefont {M.~S.}\ \bibnamefont {Zubairy}},\
  }\href@noop {} {\bibfield  {journal} {\bibinfo  {journal} {Phys. Rev. A}\
  }\textbf {\bibinfo {volume} {61}},\ \bibinfo {pages} {023609} (\bibinfo
  {year} {2000}{\natexlab{a}})}\BibitemShut {NoStop}%
\bibitem [{\citenamefont {Scully}\ and\ \citenamefont
  {Svidzinsky}(2006)}]{ScullySvidzinsky2006}%
  \BibitemOpen
  \bibfield  {author} {\bibinfo {author} {\bibfnamefont {M.~O.}\ \bibnamefont
  {Scully}}\ and\ \bibinfo {author} {\bibfnamefont {A.~A.}\ \bibnamefont
  {Svidzinsky}},\ }\href@noop {} {\bibfield  {journal} {\bibinfo  {journal} {J.
  Mod. Opt.}\ }\textbf {\bibinfo {volume} {53}},\ \bibinfo {pages} {2399}
  (\bibinfo {year} {2006})}\BibitemShut {NoStop}%
\bibitem [{\citenamefont {Svidzinsky}\ and\ \citenamefont
  {Scully}(2006)}]{SvidzinskyScully2006}%
  \BibitemOpen
  \bibfield  {author} {\bibinfo {author} {\bibfnamefont {A.~A.}\ \bibnamefont
  {Svidzinsky}}\ and\ \bibinfo {author} {\bibfnamefont {M.~O.}\ \bibnamefont
  {Scully}},\ }\href@noop {} {\bibfield  {journal} {\bibinfo  {journal} {Phys.
  Rev. Lett.}\ }\textbf {\bibinfo {volume} {97}},\ \bibinfo {pages} {190402}
  (\bibinfo {year} {2006})}\BibitemShut {NoStop}%
\bibitem [{\citenamefont {Svidzinsky}\ and\ \citenamefont
  {Scully}(2010)}]{SvidzinskyScully2010}%
  \BibitemOpen
  \bibfield  {author} {\bibinfo {author} {\bibfnamefont {A.~A.}\ \bibnamefont
  {Svidzinsky}}\ and\ \bibinfo {author} {\bibfnamefont {M.~O.}\ \bibnamefont
  {Scully}},\ }\href@noop {} {\bibfield  {journal} {\bibinfo  {journal} {Phys.
  Rev. A}\ }\textbf {\bibinfo {volume} {82}},\ \bibinfo {pages} {063630}
  (\bibinfo {year} {2010})}\BibitemShut {NoStop}%
\bibitem [{\citenamefont {Band}\ and\ \citenamefont
  {Heller}(1988)}]{BandHeller1988}%
  \BibitemOpen
  \bibfield  {author} {\bibinfo {author} {\bibfnamefont {Y.~B.}\ \bibnamefont
  {Band}}\ and\ \bibinfo {author} {\bibfnamefont {D.~F.}\ \bibnamefont
  {Heller}},\ }\href@noop {} {\bibfield  {journal} {\bibinfo  {journal} {Phys.
  Rev. A}\ }\textbf {\bibinfo {volume} {38}},\ \bibinfo {pages} {1885}
  (\bibinfo {year} {1988})}\BibitemShut {NoStop}%
\bibitem [{\citenamefont {Stepanov}(1957)}]{Stepanov1957}%
  \BibitemOpen
  \bibfield  {author} {\bibinfo {author} {\bibfnamefont {B.~I.}\ \bibnamefont
  {Stepanov}},\ }\href@noop {} {\bibfield  {journal} {\bibinfo  {journal}
  {Dokl. Akad. Nauk SSSR}\ }\textbf {\bibinfo {volume} {112}},\ \bibinfo
  {pages} {839} (\bibinfo {year} {1957})},\ \translation{Sov. Phys. Dokl.
  \textbf{2}, 81 (1957)}\BibitemShut {NoStop}%
\bibitem [{\citenamefont {Glauber}(1963{\natexlab{a}})}]{Glauber1963}%
  \BibitemOpen
  \bibfield  {author} {\bibinfo {author} {\bibfnamefont {R.~J.}\ \bibnamefont
  {Glauber}},\ }\href@noop {} {\bibfield  {journal} {\bibinfo  {journal} {Phys.
  Rev.}\ }\textbf {\bibinfo {volume} {130}},\ \bibinfo {pages} {2529} (\bibinfo
  {year} {1963}{\natexlab{a}})}\BibitemShut {NoStop}%
\bibitem [{\citenamefont {Glauber}(1963{\natexlab{b}})}]{Glauber1963b}%
  \BibitemOpen
  \bibfield  {author} {\bibinfo {author} {\bibfnamefont {R.~J.}\ \bibnamefont
  {Glauber}},\ }\href@noop {} {\bibfield  {journal} {\bibinfo  {journal} {Phys.
  Rev.}\ }\textbf {\bibinfo {volume} {131}},\ \bibinfo {pages} {2766} (\bibinfo
  {year} {1963}{\natexlab{b}})}\BibitemShut {NoStop}%
\bibitem [{\citenamefont {Kolobov}(1999)}]{Kolobov1999}%
  \BibitemOpen
  \bibfield  {author} {\bibinfo {author} {\bibfnamefont {M.~I.}\ \bibnamefont
  {Kolobov}},\ }\href@noop {} {\bibfield  {journal} {\bibinfo  {journal} {Rev.
  Mod. Phys.}\ }\textbf {\bibinfo {volume} {71}},\ \bibinfo {pages} {1539}
  (\bibinfo {year} {1999})}\BibitemShut {NoStop}%
\bibitem [{\citenamefont {Loudon}(2000)}]{Loudon2000}%
  \BibitemOpen
  \bibfield  {author} {\bibinfo {author} {\bibfnamefont {R.}~\bibnamefont
  {Loudon}},\ }\href@noop {} {\emph {\bibinfo {title} {The Quantum Theory of
  Light}}},\ \bibinfo {edition} {3rd}\ ed.\ (\bibinfo  {publisher} {Oxford
  University Press},\ \bibinfo {address} {New York},\ \bibinfo {year}
  {2000})\BibitemShut {NoStop}%
\bibitem [{\citenamefont {Bagnato}\ and\ \citenamefont
  {Kleppner}(1991)}]{BagnatoKleppner1991}%
  \BibitemOpen
  \bibfield  {author} {\bibinfo {author} {\bibfnamefont {V.}~\bibnamefont
  {Bagnato}}\ and\ \bibinfo {author} {\bibfnamefont {D.}~\bibnamefont
  {Kleppner}},\ }\href@noop {} {\bibfield  {journal} {\bibinfo  {journal}
  {Phys. Rev. A}\ }\textbf {\bibinfo {volume} {44}},\ \bibinfo {pages} {7439}
  (\bibinfo {year} {1991})}\BibitemShut {NoStop}%
\bibitem [{\citenamefont {Haugset}\ \emph {et~al.}(1997)\citenamefont
  {Haugset}, \citenamefont {Haugerud},\ and\ \citenamefont
  {Andersen}}]{HaugsetHaugerudAndersen1997}%
  \BibitemOpen
  \bibfield  {author} {\bibinfo {author} {\bibfnamefont {T.}~\bibnamefont
  {Haugset}}, \bibinfo {author} {\bibfnamefont {H.}~\bibnamefont {Haugerud}}, \
  and\ \bibinfo {author} {\bibfnamefont {J.~O.}\ \bibnamefont {Andersen}},\
  }\href@noop {} {\bibfield  {journal} {\bibinfo  {journal} {Phys. Rev. A}\
  }\textbf {\bibinfo {volume} {55}},\ \bibinfo {pages} {2922} (\bibinfo {year}
  {1997})}\BibitemShut {NoStop}%
\bibitem [{\citenamefont {Mullin}(1997)}]{Mullin1997}%
  \BibitemOpen
  \bibfield  {author} {\bibinfo {author} {\bibfnamefont {W.~J.}\ \bibnamefont
  {Mullin}},\ }\href@noop {} {\bibfield  {journal} {\bibinfo  {journal} {J. Low
  Temp. Phys.}\ }\textbf {\bibinfo {volume} {106}},\ \bibinfo {pages} {615}
  (\bibinfo {year} {1997})}\BibitemShut {NoStop}%
\bibitem [{\citenamefont {Weiss}\ and\ \citenamefont
  {Wilkens}(1997)}]{WeissWilkens1997}%
  \BibitemOpen
  \bibfield  {author} {\bibinfo {author} {\bibfnamefont {C.}~\bibnamefont
  {Weiss}}\ and\ \bibinfo {author} {\bibfnamefont {M.}~\bibnamefont
  {Wilkens}},\ }\href@noop {} {\bibfield  {journal} {\bibinfo  {journal} {Opt.
  Express}\ }\textbf {\bibinfo {volume} {1}},\ \bibinfo {pages} {272} (\bibinfo
  {year} {1997})}\BibitemShut {NoStop}%
\bibitem [{\citenamefont {Pitaevskii}\ and\ \citenamefont
  {Stringari}(2003)}]{PitaevskiiStringari2003}%
  \BibitemOpen
  \bibfield  {author} {\bibinfo {author} {\bibfnamefont {L.}~\bibnamefont
  {Pitaevskii}}\ and\ \bibinfo {author} {\bibfnamefont {S.}~\bibnamefont
  {Stringari}},\ }\href@noop {} {\emph {\bibinfo {title} {Bose-Einstein
  Condensation}}}\ (\bibinfo  {publisher} {Clarendon},\ \bibinfo {address}
  {Oxford},\ \bibinfo {year} {2003})\BibitemShut {NoStop}%
\bibitem [{\citenamefont {Kocharovsky}\ \emph {et~al.}(2006)\citenamefont
  {Kocharovsky}, \citenamefont {\relax{Vl}. V.~Kocharovsky}, \citenamefont
  {Holthaus}, \citenamefont {Ooi}, \citenamefont {Svidzinsky}, \citenamefont
  {Ketterle},\ and\ \citenamefont {Scully}}]{KocharovskyEtal2006}%
  \BibitemOpen
  \bibfield  {author} {\bibinfo {author} {\bibfnamefont {V.~V.}\ \bibnamefont
  {Kocharovsky}}, \bibinfo {author} {\bibnamefont {\relax{Vl}.
  V.~Kocharovsky}}, \bibinfo {author} {\bibfnamefont {M.}~\bibnamefont
  {Holthaus}}, \bibinfo {author} {\bibfnamefont {C.~H.~R.}\ \bibnamefont
  {Ooi}}, \bibinfo {author} {\bibfnamefont {A.}~\bibnamefont {Svidzinsky}},
  \bibinfo {author} {\bibfnamefont {W.}~\bibnamefont {Ketterle}}, \ and\
  \bibinfo {author} {\bibfnamefont {M.~O.}\ \bibnamefont {Scully}},\
  }\href@noop {} {\bibfield  {journal} {\bibinfo  {journal} {Adv. At. Mol. Opt.
  Phys.}\ }\textbf {\bibinfo {volume} {53}},\ \bibinfo {pages} {291} (\bibinfo
  {year} {2006})}\BibitemShut {NoStop}%
\bibitem [{\citenamefont {Kocharovsky}\ and\ \citenamefont {\relax{Vl}.
  V.~Kocharovsky}(2010)}]{KocharovskyKocharovsky2010}%
  \BibitemOpen
  \bibfield  {author} {\bibinfo {author} {\bibfnamefont {V.~V.}\ \bibnamefont
  {Kocharovsky}}\ and\ \bibinfo {author} {\bibnamefont {\relax{Vl}.
  V.~Kocharovsky}},\ }\href@noop {} {\bibfield  {journal} {\bibinfo  {journal}
  {Phys. Rev. A}\ }\textbf {\bibinfo {volume} {81}},\ \bibinfo {pages} {033615}
  (\bibinfo {year} {2010})}\BibitemShut {NoStop}%
\bibitem [{\citenamefont {Shiryaev}(1996)}]{Shiryaev1996}%
  \BibitemOpen
  \bibfield  {author} {\bibinfo {author} {\bibfnamefont {A.~N.}\ \bibnamefont
  {Shiryaev}},\ }\href@noop {} {\emph {\bibinfo {title} {Probability}}},\
  \bibinfo {edition} {2nd}\ ed.,\ \bibinfo {series} {Graduate Texts in
  Mathematics}, Vol.~\bibinfo {volume} {95}\ (\bibinfo  {publisher}
  {Springer},\ \bibinfo {address} {New York},\ \bibinfo {year}
  {1996})\BibitemShut {NoStop}%
\bibitem [{\citenamefont {Mandel}(1979)}]{Mandel1979}%
  \BibitemOpen
  \bibfield  {author} {\bibinfo {author} {\bibfnamefont {L.}~\bibnamefont
  {Mandel}},\ }\href@noop {} {\bibfield  {journal} {\bibinfo  {journal} {Opt.
  Lett.}\ }\textbf {\bibinfo {volume} {4}},\ \bibinfo {pages} {205} (\bibinfo
  {year} {1979})}\BibitemShut {NoStop}%
\bibitem [{\citenamefont {Barnett}\ \emph {et~al.}(1998)\citenamefont
  {Barnett}, \citenamefont {Jeffers}, \citenamefont {Gatti},\ and\
  \citenamefont {Loudon}}]{BarnettEtal1998}%
  \BibitemOpen
  \bibfield  {author} {\bibinfo {author} {\bibfnamefont {S.~M.}\ \bibnamefont
  {Barnett}}, \bibinfo {author} {\bibfnamefont {J.}~\bibnamefont {Jeffers}},
  \bibinfo {author} {\bibfnamefont {A.}~\bibnamefont {Gatti}}, \ and\ \bibinfo
  {author} {\bibfnamefont {R.}~\bibnamefont {Loudon}},\ }\href@noop {}
  {\bibfield  {journal} {\bibinfo  {journal} {Phys. Rev. A}\ }\textbf {\bibinfo
  {volume} {57}},\ \bibinfo {pages} {2134} (\bibinfo {year}
  {1998})}\BibitemShut {NoStop}%
\bibitem [{\citenamefont {Foster}\ \emph {et~al.}(2000)\citenamefont {Foster},
  \citenamefont {Mielke},\ and\ \citenamefont
  {Orozco}}]{FosterMielkeOrozco2000}%
  \BibitemOpen
  \bibfield  {author} {\bibinfo {author} {\bibfnamefont {G.~T.}\ \bibnamefont
  {Foster}}, \bibinfo {author} {\bibfnamefont {S.~L.}\ \bibnamefont {Mielke}},
  \ and\ \bibinfo {author} {\bibfnamefont {L.~A.}\ \bibnamefont {Orozco}},\
  }\href@noop {} {\bibfield  {journal} {\bibinfo  {journal} {Phys. Rev. A}\
  }\textbf {\bibinfo {volume} {61}},\ \bibinfo {pages} {053821} (\bibinfo
  {year} {2000})}\BibitemShut {NoStop}%
\bibitem [{\citenamefont {Kocharovsky}\ \emph
  {et~al.}(2000{\natexlab{b}})\citenamefont {Kocharovsky}, \citenamefont
  {\relax{Vl}. V.~Kocharovsky},\ and\ \citenamefont
  {Scully}}]{KocharovskyKocharovskyScully2000}%
  \BibitemOpen
  \bibfield  {author} {\bibinfo {author} {\bibfnamefont {V.~V.}\ \bibnamefont
  {Kocharovsky}}, \bibinfo {author} {\bibnamefont {\relax{Vl}.
  V.~Kocharovsky}}, \ and\ \bibinfo {author} {\bibfnamefont {M.~O.}\
  \bibnamefont {Scully}},\ }\href@noop {} {\bibfield  {journal} {\bibinfo
  {journal} {Phys. Rev. A}\ }\textbf {\bibinfo {volume} {61}},\ \bibinfo
  {pages} {053606} (\bibinfo {year} {2000}{\natexlab{b}})}\BibitemShut
  {NoStop}%
\end{thebibliography}
\end{document}